\documentclass[prd,reprint,superscriptaddress,nofootinbib,nobalancelastpage]{revtex4-1}
\usepackage{amsmath}
\usepackage{amssymb}
\usepackage{amsfonts}
\usepackage{graphicx}
\usepackage[caption=false]{subfig}
\usepackage{enumitem}
\usepackage[usenames,dvipsnames]{color}
\usepackage{verbatim}
\usepackage[percent]{overpic}
\usepackage{bm}
\usepackage{rotating}
\usepackage{hyperref}
\usepackage{pbox}
\usepackage{array}
\usepackage{physics}
\usepackage{mathtools}
\usepackage{leftidx}
\usepackage{lmodern}
\usepackage[normalem]{ulem}
\usepackage{cases}
\usepackage{braket}
\usepackage{tensor}
\usepackage{mathrsfs}
\usepackage{float}
\usepackage{tikz}
\usetikzlibrary{arrows.meta,decorations.pathmorphing,math}

\newcommand{\ii}{\mathrm{i}}

\newcommand{\pd}{\partial}
\newcommand{\R}{\mathbb R}

\newcommand{\Z}{\mathbb Z}
\newcommand{\rr}[1]{\left(#1\right)}

\newcommand{\sx}{\mathsf{x}}
\newcommand{\be}{\begin{equation}\begin{split}}
\newcommand{\ee}{\end{split}\end{equation}}
\newcommand{\bx}{\bm{x}}

\newcommand{\bk}{\bm{k}}

\newcommand{\mx}[1]{\begin{pmatrix}#1\end{pmatrix}}
\newcommand{\field}{\hat\rho_{\hat\phi}}
\newcommand{\allA}{\mathcal{A}}

\begin{document}


	\title{Zero mode suppression of superluminal signals in light-matter
    interactions}

	\author{Erickson Tjoa}
	\email{e2tjoa@uwaterloo.ca}
	\affiliation{Department of Physics and Astronomy, University of Waterloo, Waterloo, Ontario, N2L 3G1, Canada}
	\affiliation{Institute for Quantum Computing, University of Waterloo, Waterloo, Ontario, N2L 3G1, Canada}

	\author{Eduardo Mart\'{i}n-Mart\'{i}nez}
	\email{emartinmartinez@uwaterloo.ca}
	\affiliation{Department of Applied Mathematics, University of Waterloo, Waterloo, Ontario, N2L 3G1, Canada}
	\affiliation{Institute for Quantum Computing, University of Waterloo, Waterloo, Ontario, N2L 3G1, Canada}
	\affiliation{Perimeter Institute for Theoretical Physics, 31 Caroline St N, Waterloo, Ontario, N2L 2Y5, Canada}
	
    \begin{abstract}

    We show how two Unruh-DeWitt detectors that do not couple to the zero mode of a quantum field can exchange information faster than the speed of light. We analyze the specific cases of periodic and Neumann boundary conditions in flat spacetime with arbitrary spatial dimensions, and we show that the superluminal signal strength is only polynomially suppressed with the distance to the lightcone. Therefore, in any relativistic scenario modelling the light-matter interaction where a zero mode is present, particle detectors should explicitly couple to the zero mode.
    
    \end{abstract}
    	
    \maketitle

    \clearpage
    \section{Introduction}

    The study of quantum fields via particle detectors has been a fruitful avenue of research in quantum field theory in curved spaces, quantum optics and in relativistic quantum information \cite{Unruh1979evaporation,Landulfo2009suddendeath,Lopp:2018cavity,Alsing2012review,Lin2015teleportation}. Particle detectors are non-relativistic localized quantum systems that couple locally to quantum fields obtaining information about the field state. This allows us to probe the field without invoking projective measurements \cite{Lin2014projective,Hu2012review,sorkin1956,Benincasa2014projective}. The paramount example of a particle detector is an atom coupled to the electromagnetic field. 
    Among the most successful models of particle detectors we have the Unruh-DeWitt (UDW) model, consisting of a  two-level quantum system coupled locally to a scalar field \cite{DeWitt1979}. Despite its simplicity, this model has been shown to capture the main features of the light-matter interaction \cite{Pablo2018rqo,pozas2015harvesting} and has been extensively used to study fundamental properties of quantum fields \cite{Takagi1986noise,Crispino2008review}.
    
    In solid state physics and in quantum optics, the spatial topology of the setup is something that can be given by the particular experimental setup. For example, one can have an optical fibre coiled around itself to have periodic boundary conditions in one dimension. Hence, it is natural to ask what the role of boundary conditions have in modeling the light-matter interaction, and whether assuming simpler models could lead to faster-than-light signalling between spacelike separated operators of particle detectors. For instance, it has been recently studied how factors such as the detector smearing, rotating-wave approximations or the introduction of UV regularization have implications on causality in particle detector models \cite{Causality2015Eduardo}. 
    
    It is known that in $(1+1)$ dimensional flat spacetime, a scalar field subjected to periodic boundary conditions has a zero mode which contributes to a particle detector's response, the field's stress-energy tensor, and the ability for particle detectors to get entangled through the field \cite{Zeromode2014Edu,Lin2016entangleCylin}. Zero modes also appear in other contexts such as two-dimensional conformal field theories (CFTs) and in the minimal coupling of massless scalar field in certain spacetimes with nontrivial compact topology \cite{francesco1996conformal,Allen1987deSitter,Kirsten1993deSitter,Page2012deSitter,Bros2010}, where regularization schemes for the Wightman function have impacts on the zero modes.
    However, the zero mode is peculiar as compared to the regular oscillator modes since  it does not admit a Fock space representation. For this reason, it is perhaps desirable to be able to ignore or remove the zero mode from any calculation by hand. In some contexts, such as UDW model coupled via derivative coupling, its effect can indeed be made negligible at the level of detector responses in appropriate limits \cite{Zeromode2014Edu}, but in some other contexts it has significant impact on detector dynamics and entanglement \cite{Zeromode2014Edu,Louko2016fermionZM,Yazdi2017,Lin2016entangleCylin}. There are also cases when the zero mode has been excluded by assumption from a setup with periodic boundary conditions (e.g., in \cite{Robles2017thermometryQFT,Brown2013Amplification,Brenna2016antiUnruh,Hummer2016bosonfermionZM,Lorek2014tripartite}), thus it is of interest to further study the impact that the removal of the zero mode may have on the relativistic nature of the interaction, and in particular in the causality of the whole particle detector model.
    
    Here we will investigate how neglecting the zero mode of a massless scalar field can lead to faster-than-light signalling between particle detectors via violations of microcausality\footnote{Note that in the context of algebraic quantum field theory (AQFT), sometimes this is known as a version of \textit{locality} \cite{HaagKastler1964algebraic,Bros2010}.}. We will show how two particle detectors coupled locally to the field can non-negligibly communicate faster than light when the zero mode is neglected. As a consequence, we show that whenever a zero mode arises, one cannot remove it by hand and only consider the oscillator part if relativistic phenomenology is important in the setup under study. We will also show how this zero mode-induced causality violation is alleviated in higher dimensions. In this paper, we first study the causality with respect to zero mode in $(1+1)$ and $(2+1)$ dimensions and then make an argument for arbitrary dimensions. 
    
    The paper is organized as follows. In Section~\ref{sec:causalityevaluate} we briefly review the UDW model and the notion of signalling estimator, and its relation to microcausality. In Section~\ref{sec:causal1D} we study microcausality in $(1+1)$ dimensions. In Section~\ref{sec:causal2D} we study several cases in $(2+1)$ dimensions for different choices of spatial section topology. In Section~\ref{sec:discuss} we briefly discuss the general case in arbitrary dimensions. In this paper we use the natural units $c=\hbar=1$ throughout, and we use the notation for spacetime event $\sx\equiv (t,\bx)$ whenever convenient.

    \section{How do we evaluate causality?}
    \label{sec:causalityevaluate}
    When microcausality is violated, the commutator between two observables at two spacelike separated events may no longer be zero. This in turn can be used to perform faster-than-light signalling with particle detectors. To make this idea precise in an operational manner we follow \cite{Causality2015Eduardo} and we consider two observers Alice and Bob who are spacelike separated, each carrying a particle detector which can interact with the field locally. We model these detectors using a pair of Unruh-DeWitt detectors consisting of two-level quantum systems (qubits). The monopole moment of each detector in the interaction picture is given by
    \begin{equation}
        \hat\mu_\nu(\tau) = \hat \sigma_\nu^+e^{i \Omega_\nu \tau} + \hat \sigma_\nu^- e^{-i\Omega_\nu \tau}
    \end{equation}
    where $\nu=\{\textsc{a},\textsc{b}\}$ denotes Alice or Bob respectively. Here we have $\hat\sigma_\nu^+=\ket{e_\nu}\bra{g_\nu}$, $\hat\sigma_\nu^-=\ket{g_\nu}\bra{e_\nu}$ are the usual $\mathfrak{su}(2)$ ladder operators, $\ket{g_\nu},\ket{e_\nu}$ are the ground and excited states of the qubit, $\Omega_{\nu}$ is the gap of the qubit and $\tau$ is the proper time of the the qubits. Since we are in flat space, the proper time for both detectors will be the same. 
    
    The linear UDW model prescribes the following interaction between the field and a stationary detector \cite{Pablo2018rqo}, 
     \begin{equation}
        \hat H_\nu = \lambda_\nu\chi_\nu(t)\hat\mu_\nu(t)\int\dd^n\bx\, F_\nu(\bx-\bx_\nu) \hat\phi(t,\bx)\,,
    \end{equation}
    where $F(\bx-\bx_\nu)$ is the spatial smearing of the detector $\nu$, centred at $\bx_\nu$, $\chi_\nu(t)$ is the switching function of the detector, and $\lambda_\nu$ is the coupling strength. We can assume that the Hamiltonians generate translations with respect to the same time parameter for both detectors assuming they are at rest relative to each other and also relative to the lab frame where the field quantization is performed. 
    
    The full interaction Hamiltonian for the field and the two detectors is given by
    \begin{equation}
        \hat H_I(t) = \hat H_\textsc{a}(t) \otimes \openone_\textsc{b} + \openone_\textsc{a}\otimes \hat H_\textsc{b}(t)\,.
    \end{equation}
    We assume that the system is initialized in the completely uncorrelated state 
    \begin{equation}
    \begin{split}
        \hat\rho_0 
        &= \hat\rho_\textsc{a}\otimes\hat\rho_\textsc{b}\otimes \hat \rho_{\hat\phi}\,
    \end{split}
    \end{equation}
    where $\hat \rho_\phi$ is an arbitrary field state, which in the presence of a zero mode we can split as  $\hat\rho_{\hat\phi}= \hat \rho_{\text{osc}}\otimes \hat \rho_{\text{zm}}$ where $\hat \rho_{\text{osc}}$ is the state of all the modes that admit a Fock quantization and $\hat \rho_{\text{zm}}$ is the state of the zero mode. The state $\rho_\textsc{a}\otimes\rho_\textsc{b}$ is the most general product state of both detectors, which in a matrix representation in the basis 
    \begin{equation}
        \ket{e}=\mx{1\\0} \,,\hspace{0.5cm}\ket g = \mx{0\\1}
    \end{equation}
    reads
    \begin{equation}
        \begin{split}
        \hat\rho_\textsc{a}\otimes\hat\rho_\textsc{b} =
        \mx{\alpha_\textsc{a} & \beta_\textsc{a}\\
        \beta_\textsc{a}^* & 1-\alpha_\textsc{a}}\otimes\mx{\alpha_\textsc{b} & \beta_\textsc{b}\\
        \beta_\textsc{b}^* & 1-\alpha_\textsc{b}}\,.
        \end{split}
    \end{equation}
    where $\alpha_\nu \in \mathbb{R}$. 
    
   Notice that while there is an ambiguity to choose the physically meaningful state for a zero mode, all the results in this paper are independent of the state of the field, therefore we do not need to concern ourselves with discussing what would be a reasonable state for the field in general and in particular for the zero mode as long as the expectation values of the field commutators are well defined.

    The state evolves as
    \begin{equation}
        \hat\rho = \hat U\hat \rho_0\hat U^\dagger\,,
    \end{equation}
    where the time evolution operator is
    \begin{equation}
        \hat U = \mathcal{T}\exp\left[-\ii\int_{-\infty}^\infty\!\!\!\dd t\,\hat H_I(t)\right]
    \end{equation}
    and $\mathcal{T}$ denotes time-ordering. The time evolution can be found perturbatively order-by-order in the coupling strengths $\lambda_\nu$. The final state of the two-detector subsystem is then given by the reduced joint density matrix
    \begin{equation}
        \hat\rho_{\textsc{ab}} = \tr_{\hat\phi}\rr{\hat\rho} = \hat\rho_{\textsc{ab},0} + \hat\rho_{\textsc{ab}}^{(1)}+\hat\rho_{\textsc{ab}}^{(2)} + O(\lambda^3)\,,
    \end{equation}
    where the superscript $(j)$ denotes the contribution to the time-evolved density matrix of order $\lambda^j$.
    
    In order to see how signalling using particule detectors is connected to microcausality, we first note that any contributions linear in $\lambda_\textsc{a}$ or $\lambda_\textsc{b}$ are local, thus $\hat\rho^{(1)}$ cannot contribute to signalling. The signalling part of the detectors' density matrix $\hat\rho_{\textsc{ab}}$ must be of second order in the product of coupling strengths $\lambda_\textsc{a}
    \lambda_{\textsc{b}}$ \cite{Causality2015Eduardo}. Therefore, we can split the second order term into three parts, namely
    \begin{equation}
        \hat\rho_{\textsc{ab}}^{(2)} = \hat\rho_{\textsc{ab},\text{signal}}^{(2)} + \hat\rho_{\textsc{a},\text{noise}}^{(2)}+\hat\rho_{\textsc{b},\text{noise}}^{(2)}\,.
    \end{equation}
    The last two terms are of the order $\lambda_{\textsc{a}}^2$ and $\lambda_{\textsc{b}}^2$ respectively, hence they are local noise terms which do not contribute to signalling between the two detectors. The first term is the signalling term which is of the order $\lambda_\textsc{a}
    \lambda_{\textsc{b}}$. This can also be seen by finding the reduced state of detector B alone and only the signalling term will survive:
    \begin{equation}
        \hat\rho_{\textsc{b},\text{signal}}  = \tr_{\textsc{a}}\left(\hat \rho_{\textsc{ab},\text{signal}}^{(2)} \right)\,.
    \end{equation}
    In order to cleanly separate effect of zero mode on microcausality from the smearing and switching effects, we consider both compactly supported smearing and switching function. That is, 
    \begin{equation}
    \begin{split}
        \text{supp}[\chi_\nu(t)] &= \left[T^{\text{on}}_\nu,T^{\text{off}}_\nu\right]\,,\\
        \text{supp}[F(\bx-\bx_\nu)] &= \left[\bx_\nu-\frac{\sigma}{2},\bx_\nu+\frac{\sigma}{2}\right]\,,
    \end{split}
    \end{equation}
    where $\sigma$ is the width of the spatial smearing (i.e. an effective detector diameter).  We require that these supports do not overlap, i.e.
    \begin{equation}
        \begin{split}
        T^{\text{off}}_\textsc{a} &< T^{\text{on}}_\textsc{b}\,,\\
        \bx_\textsc{a}+\frac{\sigma}{2} &< \bx_\textsc{b}-\frac{\sigma}{2}\,.
        \end{split}
    \end{equation}
    Under these conditions, it can be shown that
    \begin{equation}
        \begin{split}
        \hat\rho_{\textsc{b},\text{signal}}^{(2)} &= 2\int_{\mathbb{R}}\dd t\int_{\mathbb{R}}\dd t'\chi_\textsc{a}(t)\chi_\textsc{b}(t')\text{Re}\rr{\beta_\textsc{a}e^{\ii\Omega_\textsc{a} t}}\mathcal{C}(\sx_\textsc{a},\sx'_\textsc{b})\\
        &\hspace{0.5cm} \mx{-2\text{Im}(\beta_{\textsc{b}}e^{\ii\Omega_\textsc{b}t'}) &  -\ii e^{-\ii\Omega_\textsc{b} t'}(1-2\alpha_\textsc{b})\\ \ii e^{-\ii\Omega_\textsc{b} t'} (1-2\alpha_\textsc{b}) & 2\text{Im}(\beta_{\textsc{b}}e^{\ii\Omega_\textsc{b}t'})}\,.
        \end{split}
    \end{equation}
    The function $\mathcal{C}(\sx_\textsc{a},\sx_\textsc{b})\equiv \mathcal{C}(t,\bx_\textsc{a},t',\bx_\textsc{b})$ in the integrand is the spatially smeared pull-back of the field commutator, as shown in detail in \cite{Causality2015Eduardo}:
    \begin{equation}
        \begin{split}
            \mathcal{C}(t,\bx_\textsc{a},t',\bx_\textsc{b})&:=\int_{\mathbb{R}^n}\!\!\! \dd^n\bx \int_{\mathbb{R}^n}\!\!\!\dd^n\bx' F_{\textsc{a}}(\bx-\bx_\textsc{a})F_\textsc{b}(\bx'-\bx_\textsc{b})\\
            &\hspace{0.5cm}\braket{[\phi(t,\bx),\phi(t',\bx')]}\,,
        \end{split}
    \end{equation}
    where $\bx_j$ are the centres of mass of the smearings of the detectors used to probe causality. To estimate the ability of A and B to perform faster-than-light signalling we analyze the  causality estimator $\mathcal{E}$ proposed in \cite{Causality2015Eduardo}, which is proportional to the signal strength of the contributions to the density matrix of detector B coming from the presence of detector A:
    \begin{equation}
        \begin{split}
        \label{eq:estimator}
            \mathcal{E}(\bx_\textsc{a},\bx_\textsc{b}):=\abs{\int_\mathbb{R}\dd t\int_\mathbb{R}\dd t'\,\chi_\textsc{a}(t)\chi_\textsc{b}(t')\mathcal{C}(t,\bx_\textsc{a},t',\bx_\textsc{b})}\,.
        \end{split}
    \end{equation}
    Furthermore, it has been shown that channel capacity, measured by a lower bound to the number of bits per unit time that can be sent from Alice to Bob, is directly related to $\mathcal{E}$ \cite{Cliche2010channel,Jonsson2015infofree}. 
  
    Notice that one can also particularize to a delta switching (that can be understood as the limit of very short time Gaussian switching when the total strength of the interaction over time is fixed, see, e.g.,  \cite{PozasJorma}). In the case of this instantaneous switching, the reduced density matrix of detector B will simply be proportional to $\mathcal{C}$, thus this function is a legitimate measure of signalling between detectors. 
    For this reason we will make use of both $\mathcal{E}$ and $\mathcal{C}$ as causality estimators in the subsequent sections. 

    \section{Causality and zero mode in (1+1) dimensions}
    \label{sec:causal1D}
    We consider massless scalar field on the Einstein cylinder with the metric \cite{birrell1984quantum}
    \begin{equation}
        \dd s^2 = -\dd t^2 + \dd x^2\,,
    \end{equation}
    where the spacetime has topology $\R\times S^1$. The topological identification is made for $x\sim x+L$, where $L$ is the circumference of the cylinder. This is the same as having a periodic cavity in $(1+1)$ dimensions, i.e., periodic boundary condition for the scalar field in Minkowski spacetime. The field operator can be decomposed into two parts,
    \begin{equation}
        \hat\phi(t,x) = \hat\phi_\text{zm}(t)+\hat\phi_{\text{osc}}(t,x)\,.
    \end{equation}
    The first term $\hat\phi_0$ is the zero mode term which is spatially constant. The second term $\hat\phi_{\text{osc}}(t,\bx)$ is the harmonic oscillator term whose mode decomposition reads
    \begin{equation}
        \begin{split}
        \hat\phi_{\text{osc}}(t,x) &= \sum_{n\neq 0} \frac{1}{\sqrt{4\pi\abs{n}}}\left[e^{-\ii\abs{k_n}t+\ii k_n x}\hat a_{n} + \text{h.c.}\right]\,,\\
        k_n &= \frac{2\pi n}{L}\,, n\in \Z\,.
        \end{split}
    \end{equation}
    The oscillator modes have a Fock vacuum $\ket{0}$ defined by $\hat a_n\ket{0}=0$ for all $n\in \mathbb{Z}$ and the usual canonical commutation relation for the ladder operators $[\hat a_j,\hat a_k^\dagger] = \delta_{jk}$.
    
    Note that the zero mode behaves as a ``free-particle": specifically, the Lagrangian only contains the kinetic part \cite{Zeromode2014Edu,francesco1996conformal}
    \begin{equation}
    \begin{split}
        \mathcal{L}_\text{zm} &= \frac{L\dot{Q}^2}{2}\,,
    \end{split}
    \end{equation}
    where $Q:=\tilde\phi_0$ is the Fourier component of the zero mode. We can think of this as an ``oscillator" with zero frequency, since after Legendre transformation the zero-mode free  Hamiltonian (after quantization) is given by:
    \begin{equation}
    \label{eq:1Dzerohamiltonian}
        \hat H_\text{zm} = \frac{\hat{P}^2}{2L}\,,\hspace{0.3cm} P = \frac{\pd \mathcal{L}_\text{zm}}{\pd \dot{Q}}\,.
    \end{equation}
    In the interaction picture, we have that
    \begin{equation}
        \hat Q(t) = \hat Q_S + \frac{\hat P_S t }{L}
    \end{equation}
    where the subscript $S$ means Schr\"odinger picture operator. The Heisenberg equation of motion then implies that
    \begin{equation}
        \hat \phi_\textsc{zm}(t) = \hat Q(t) = \hat \phi_\textsc{zm}(0) + \frac{\hat P_St}{L}\,.
    \end{equation}

    \begin{figure*}[tp]
        \centering
        \includegraphics[scale=0.8]{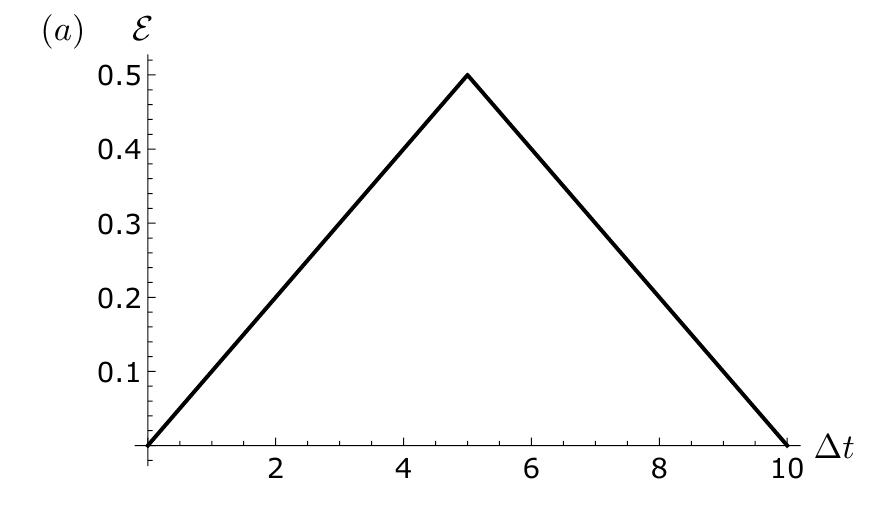}
        \includegraphics[scale=0.8]{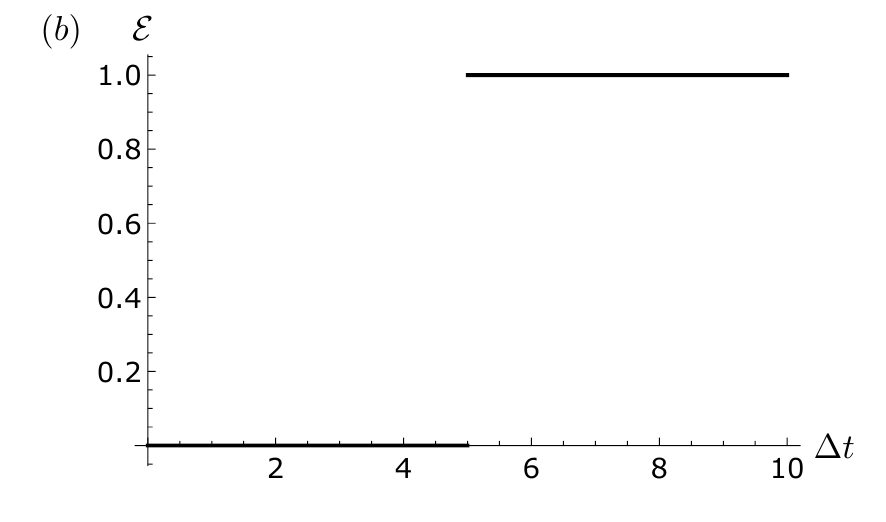}
        \caption{Causality estimator for delta switching and pointlike detector, with $\Delta x=5$ in natural units. \textbf{(a):} excluding the zero mode. The commutator does not vanish even for spacelike separated regions. \textbf{(b):} including the zero mode. Microcausality is recovered as commutator vanishes identically for $|\Delta t|<|\Delta x|$.}
        \label{fig:delta_causal}
    \end{figure*}
    
    The field commutator can be written as the sum of the commutator for the oscillator modes and the zero mode, 
    \begin{equation}
        [\hat\phi(\sx),\hat\phi(\sx')] = [\hat\phi_\text{zm}(t),\hat\phi_\text{zm}(t')]+[\hat\phi_\text{osc}(\sx),\hat\phi_\text{osc}(\sx')]\,.    
        \label{eq:fullcommutatoranyD}
    \end{equation}
    The oscillator contribution to the commutator is given by (see Appendix~\ref{appendix: commutators})
    \begin{align}
        [\hat\phi_{\text{osc}}(\sx),\hat\phi_{\text{osc}}(\sx')]
        &=-\frac{1}{4\pi}\log\rr{1-e^{-\frac{2\ii\pi(\Delta u-\ii\epsilon)}{L}}} \notag\\
        &\hspace{0.3cm}-\frac{1}{4\pi}\log\rr{1-e^{-\frac{2\ii\pi(\Delta v-\ii\epsilon)}{L}}}\notag\\
        &\hspace{0.3cm}+\frac{1}{4\pi}\log\rr{1-e^{\frac{2\ii\pi(\Delta u-\ii\epsilon)}{L}}}\notag\\
        &\hspace{0.3cm}+\frac{1}{4\pi}\log\rr{1-e^{\frac{2\ii\pi(\Delta v-\ii\epsilon)}{L}}}\,,
        \label{eq:osccommutator1D}
    \end{align}
    where $u = t-x$ and $ v=t+x$ are the double null coordinates in Minkowski space and $\Delta u = u-u'$ and $\Delta v=v-v'$. The commutator due to the zero mode reads \cite{Zeromode2014Edu}
    \begin{equation}
        [\hat\phi_\text{zm}(\sx),\hat\phi_\text{zm}(
        \sx')] = -\frac{\ii\Delta t}{L}\,,\hspace{0.5cm} \Delta t = t-t'\,.
        \label{eq:1Dzero}
    \end{equation}
    
    Let us now check the causality estimators in $(1+1)$ dimensions. The simplest case is when we take pointlike detectors and instantaneous switching, which reduces the estimators $\mathcal{E}$ to be proportional to $\mathcal{C}$. Note that even if we do not know the ground state for the zero mode, commutator is a $c$-number so the causality estimator is state-independent.

    In general, for a pointlike detector in arbitrary dimensions and delta switching, one can run into UV-divergent detector reduced density matrix. However, the causality estimator is UV-safe and does not have such problems even in the limiting cases where UV-divergences may appear \cite{Blasco2015Huygens,Blasco2016broadcast}. Note as well that we can always avoid this problem by not taking both limits (infinitely fast switching and pointlike smearing) simultaneously.
    
    In Figure~\ref{fig:delta_causal}(a) we show the causality estimator \eqref{eq:estimator} for a delta switching and pointlike detectors for $L=10$, $\Delta x = 5$ (that is, the separation between the detectors so that $\Delta t<5$ corresponds to spacelike separation). The figure demonstrates causality violation when one removes the zero mode contribution.  The causality violation coming from ignoring the zero mode is very strong as it can be seen in the figure. The decay of the signalling contributions (thus the decay of the superluminal channel capacity between Alice and Bob) decays only linearly with the distance to the light cone.
    
   When we plot the whole commutator including the zero mode in Figure \ref{fig:delta_causal}(b) we recover the full causal behaviour: the commutator vanishes in the spatial separation domain $\Delta t<5$.   We should also note that in $(1+1)$ dimensions we have a violation of \textit{strong Huygens' principle} \cite{Blasco2015Huygens,Blasco2016broadcast,Jonsson2015infofree,Sonego1992huygens,Valerio1999tails}, i.e. the support of the commutator is on the whole timelike region bounded by the light cone, and in fact it is constant inside the lightcone.

\begin{figure*}[tp]
        \includegraphics[scale=0.9]{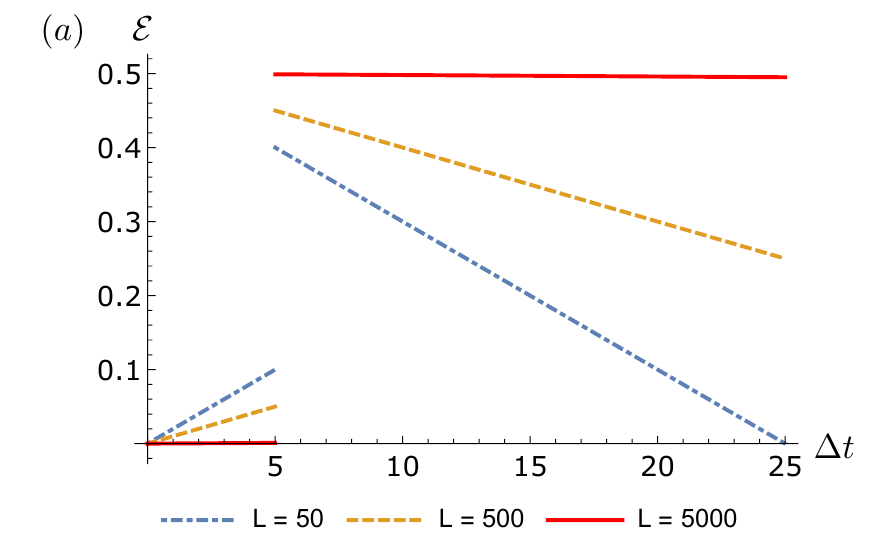} 
        \includegraphics[scale=0.9]{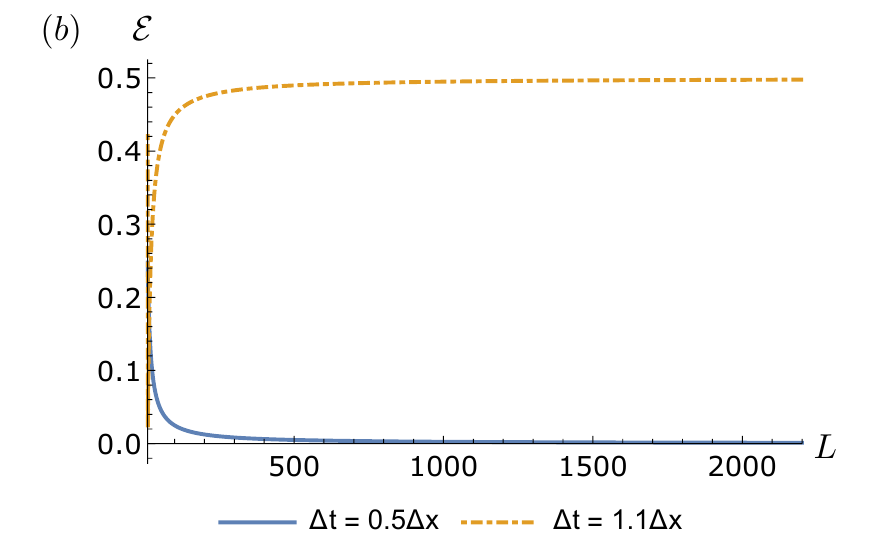}
        \caption{\textbf{(a):} Causality estimator of the purely oscillator part as a function of time gap between detector switching times $\Delta t$ for several choices of $L$. For large $L$ microcausality is approximately recovered. \textbf{(b):} Causality estimator of the purely oscillator part as a function of $L$. We see that the causality estimator falls quickly with increasing $L$ when detectors are spacelike separated and quickly approaches constant value when timelike separated. }
        \label{fig:delta_causal2}
    \end{figure*}

   Note that the zero mode contribution is inversely proportional to $L$. As one may have expected, the oscillator mode contribution to the commutator dominates at large $L$ and also uniformly becomes microcausal for large $L$, as shown in  Figure~\ref{fig:delta_causal2}(a). In other words, if the cavity is large, the  causality violation is small when one ignores zero mode contribution of the quantum field. Consequently, the usual ``toroidal" quantization used e.g. in \cite{birrell1984quantum} where one puts a field in a torus and take $L\to \infty$ to reproduce free space quantization does not suffer causality violation because of the limit taken.  However, one has to be careful since the superluminal signaling decays only linearly with the length $L$, hence the faster-than-light signalling will not be strongly suppressed. This is illustrated in  Figure~\ref{fig:delta_causal2}(b).
   
\begin{figure}[tp]
        \includegraphics[scale=1]{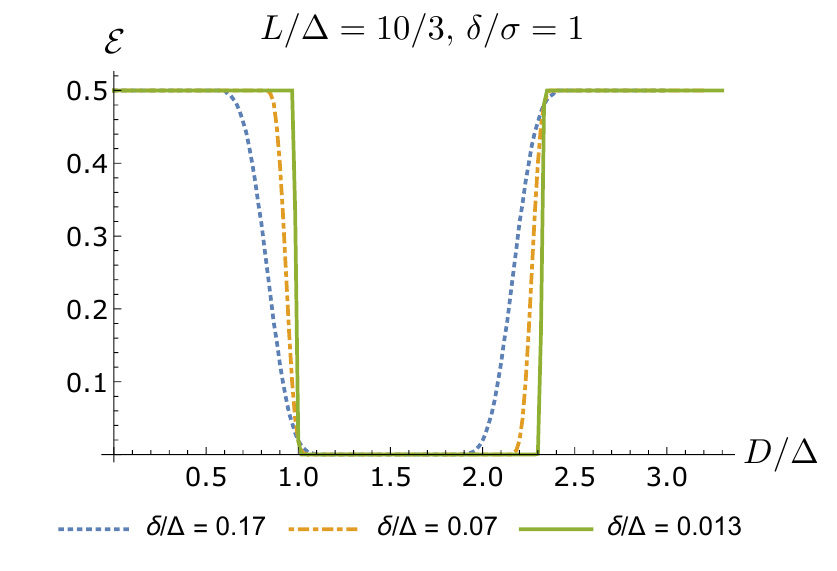}
        \caption{Causality estimator $\mathcal{E}$ as a function of outer distance of the finite-sized detectors $D$ for various switching duration $\delta$ and size of detector $\sigma$. The time gap between the switch-off of detector A and switch-on of detector B is denoted $\Delta$.}
        \label{fig:spatial_causal1}
    \end{figure}

    In a more general setting, we could consider the presence of compactly supported spatial smearing and switching functions. In Figure~\ref{fig:spatial_causal1} we show the causality estimator $\mathcal{E}$ when we include the zero mode for various choices of detector size $\sigma$ and duration of switching $\delta$ for each detector. In this plot we used the hard-sphere smearing and finite Heaviside switching of the form
    \begin{equation}
        \begin{split}
        \chi_\nu(t) &= \begin{cases}1/\delta \hspace{0.5cm} t\in \left[T^{\text{on}}_\nu,T^{\text{off}}_\nu\right]\\
        0 \hspace{0.9cm} \text{otherwise}\end{cases}\,,\\
        F_\nu(x-x_\nu) &= \begin{cases}1/\sigma \hspace{0.5cm} x \in \left[\bx_\nu-\frac{\sigma}{2},\bx_\nu+\frac{\sigma}{2}\right]\\
        0 \hspace{0.9cm} \text{otherwise}\end{cases}
        \end{split}
    \end{equation}
    where $\delta := T^{\text{on}}_\nu-T^{\text{off}}_\nu$ is the duration of the switching which we set to be equal for both detectors and $\sigma$ is the finite size of both detectors.
    We also fix the time gap between the two detector's switch-on/off times $\Delta:=T^{\text{on}}_\textsc{b}-T^{\text{off}}_\textsc{a}$ and $D$ is the surface-to-surface distance between both detectors. We choose $\delta/\sigma=1$ in all cases but we decrease the value of $\delta/\Delta$, which amounts to shorter switching duration \textit{and} smaller detector size. Indeed, we see that the causality estimator approaches the delta switching and pointlike limit. These results also indicate that causality estimator $\mathcal{E}$ is largely independent of the type of switching or smearing functions and mainly dependent on their durations/lengths. Therefore, to discuss causality violations in detector signalling for higher dimensional cases it suffices to focus on the pointlike and fast-switching limits for $\mathcal{E}$.

    We also note here that when we impose Neumann boundary condition instead of periodic boundary condition, it will also yield a zero mode. In this case, the spacetime still has the same metric as Minkowski space, but now we consider homogeneous Neumann boundary condition
    \begin{equation}
        \left.\frac{\pd \hat\phi}{\pd x}\right|_{x=0} = \left.\frac{\pd \hat\phi}{\pd x}\right|_{x=L} = 0\,. 
    \end{equation}
    The eigenfunctions now take the form
    \begin{equation}
        u_{n}(t,x) = N_n\cos\frac{n\pi x}{L}e^{-\ii |k_n|t}\,, \hspace{0.5cm} n\in \mathbb{N}\cup 0\,.
    \end{equation}
    The spatially constant solution $u_0(t,x)$ corresponds to the zero mode. Therefore, the Klein-Gordon inner product only works for $n\in \mathbb{N}$ which gives $N_n = 1/\sqrt{n\pi}$ and the zero mode $u_n(t,x)$ has to be treated separately. The oscillator part of the commutator now reads (see Appendix~\ref{appendix: commutators})
    \begin{widetext}
    \begin{align}
         &[\hat\phi_{\text{osc}}(\sx)\hat\phi_{\text{osc}}(\sx')]\notag\\
         &=\frac{1}{4\pi}\left[
         \log \left(1-e^{\frac{\ii \pi(u-v'-\ii\epsilon)}{L}}\right)+
         \log \left(1-e^{\frac{\ii \pi(\Delta v-\ii\epsilon)}{L}}\right)+
         \log \left(1-e^{\frac{\ii \pi(\Delta u-\ii\epsilon)}{L}}\right)+
         \log \left(1-e^{\frac{\ii \pi(v-u'-\ii\epsilon)}{L}}\right)\right.\notag\\
         &\hspace{0.75cm}\left.
         -\log \left(1-e^{-\frac{i \pi (u-v'-\ii\epsilon)}{L}}\right)
         -\log \left(1-e^{-\frac{i \pi (\Delta v-\ii\epsilon)}{L}}\right)
         -\log \left(1-e^{-\frac{i \pi (\Delta u-\ii\epsilon)}{L}}\right)
         -\log \left(1-e^{-\frac{i \pi (v-u'-\ii\epsilon)}{L}}\right)
         \right]\,.
         \label{eq:osccommutator1Dneumann}
    \end{align}
    \end{widetext}
    
    This commutator differs from the case for periodic boundary conditions shown in Eq.~\eqref{eq:osccommutator1D} by a factor of $2$ in the momentum $k_n$ and the fact that the commutator is no longer translation-invariant. The zero mode commutator remains the same as before. The estimator $\mathcal{E}$ for Neumann boundary condition will be similar to the periodic boundary case shown previously in Figure~\ref{fig:delta_causal}, thus we do not repeat the plot for Neumann boundary conditions.
    
    Last but not least, there is an interesting observation we can make regarding the expressions for the commutators: if we invoke the identity
    \begin{align}
        &\log\frac{(1-e^{\ii\phi})(1-e^{\ii\psi})}{(1-e^{-\ii\phi})(1-e^{-\ii\psi})} \notag\\
        &\hspace{1.5cm}\equiv \log e^{\ii(\phi+\psi)} \notag\\
        &\hspace{1.5cm}= i(\phi+\psi) + 2\pi \ii n\,,\hspace{0.5cm} n\in \mathbb{Z}\,,
        \label{eq:logidentity}
    \end{align}
    the expression seems to simplify considerably. However, one has to be careful with the branch cuts of the logarithm when applying this simplification. When the detectors are spacelike separated we do not cross the branch cut of the of the logarithm when taking its principal branch $(n=0)$. In that case, taking the principal branch of the logarithm, the oscillator contribution $[\hat\phi_{\text{osc}}(\sx),\hat\phi_{\text{osc}}(\sx')]$ for \textit{both} the periodic and Neumann boundary conditions in Eq.~\eqref{eq:osccommutator1D} and Eq.~\eqref{eq:osccommutator1Dneumann} appear to simplify further into
    \begin{align}
         [\hat\phi_{\text{osc}}(\sx),\hat\phi_{\text{osc}}(\sx')] &= 
         \frac{1}{4\pi}\log\left[e^{\frac{4\pi\ii(\Delta t-\ii\epsilon)}{L}}\right] \notag\\
         &= \frac{\ii\Delta t}{L}+\frac{\epsilon}{L}\,.
         \label{eq:oscillator1Dbranch}
    \end{align}
    Consequently, by adding Eq.~\eqref{eq:osccommutator1D} or Eq.~\eqref{eq:osccommutator1Dneumann} to the zero mode commutator $ [\hat\phi_{\text{zm}}(\sx),\hat\phi_{\text{zm}}(\sx')]$, followed by the limit $\epsilon\to 0$, we get the following simple result for for spacelike separated $\sx,\sx'$:
    \begin{align}
        [\hat\phi(\sx),\hat\phi(\sx')] = 0\,,
    \end{align}
    as it should be if microcausality is not violated.
    
    However, there are some subtleties associated with the above simplifications. For one, the identity seems to hide the role of spatial separation $\Delta x$ because only $\Delta t$ appears in the expression for $[\hat\phi_{\text{osc}}(\sx),\hat\phi_{\text{osc}}(\sx')]$ in Eq.~\eqref{eq:oscillator1Dbranch}. It turns out that depending on the values of $t,t',x,x'$, we may cross branch cuts and the terms in the commutator may refer to different branches of the logarithm. More specifically, from Eq.~\eqref{eq:oscillator1Dbranch} we can deduce that the simplification holds for spacelike separated $\sx,\sx'$ whenever the spatial difference $\Delta x$ satisfies
    \begin{align}
        \frac{|\Delta x|}{L} \equiv \frac{|x-x'|}{L} \geq \frac{1}{4}\,,
    \end{align}
    otherwise we will need to use the full expression given in Eq.~\eqref{eq:osccommutator1D}. For arbitrary separation, using Eq.~\eqref{eq:logidentity} the simplification will read
    \begin{align}
        [\hat\phi(\sx),\hat\phi(\sx')] = \frac{n}{2}\,,\hspace{0.5cm} n\in \mathbb{Z}\,.
    \end{align}
    Here $n$ refers to different branches of the full simplified logarithm in Eq.~\eqref{eq:oscillator1Dbranch} which depends on $t,t',x,x'$ in nontrivial manner. The timelike separated case as shown in Figure~\ref{fig:delta_causal}(b) is in fact the $n=2$ branch. For arbitrary values of $x,x',t,t'$ the value of $n$ will depend on how many logarithms in the sums in Eq.~\eqref{eq:osccommutator1D}  and \eqref{eq:osccommutator1Dneumann} cross branch cuts for the value of the parameters. The consequent piecewise simplification of the zero-mode commutator would in general be cumbersome so we only included it in detail for the spacelike case which is the one we focus on to study causality.


    \section{Causality and zero mode in (2+1) dimensions}
    \label{sec:causal2D}
    
    In $(1+1)$ dimensions, we showed that both periodic and (homogeneous) Neumann boundary conditions have zero modes which lead to causality violations when they are removed unjudiciously. Both boundary conditions are essentially unique since there is only one way to implement them. For example, in $(1+1)$ dimensions there is a unique spatial topology corresponding to periodic boundary conditions, namely $S^1$. Similarly, there is only one possible homogeneous Neumann boundary condition, namely spatial derivatives at both ends are set to zero. In higher dimensions, there are more possibilities due to more freedom in imposing the boundary conditions. For instance, homogeneous Neumann boundary conditions can be implemented for various boundary shapes, and one can impose periodic boundary condition on one dimension and, e.g., Dirichlet boundary condition on the remaining spatial dimensions.

    \subsection{Annular boundary condition: $\Sigma=I\times S^1$}
    The simplest case we consider will involve a two-dimensional `annular' cavity, where the spatial topology is $I\times S^1$ where $I\subset\R$ is a compact interval. This is equivalent to taking the massless scalar field in Minkowski spacetime but impose Dirichlet boundary conditions in one direction and periodic boundary conditions in another. If we let $x$ to be the coordinate with the periodic boundary condition and $y$ the coordinate with the Dirichlet boundary condition, we have
    \begin{equation}
        \begin{split}
            \hat\phi(t,x,y) &= \hat\phi(t,x+L_1,y)\,,\\
            \hat\phi(t,x,0) &= \hat\phi(t,x,L_2) = 0\,.
        \end{split}
    \end{equation}
    For convenience we consider the case with $L_1=L_2=L$. The positive eigenmodes with respect to Minkowski timelike Killing vector for this case is given by
    \begin{equation}
    \begin{split}
        u_{nl}(t,x,y) &= N_{nl}e^{-\ii\abs{\bk_{nl}}t}\exp\frac{2\ii\pi n  x}{L}\sin\frac{l\pi y}{L}\,,\\
        \abs{\bk_{nl}} &= \sqrt{\rr{\frac{2\pi n}{L}}^2+\rr{\frac{\pi l}{L}}^2} \,, n\in \mathbb{Z},l\in \mathbb{N}\,,
    \end{split}
    \end{equation}
    where $N_{nl}$ is a normalization constant. For clarity, we explicitly derive the normalization using the Klein-Gordon inner product:
    \begin{equation}
        \begin{split}
            \delta_{nn'}\delta_{ll'} = -\ii\int_0^L\!\! \dd x\int_0^L\!\!\dd y \,\rr{ u_{nl}\frac{\pd u^*_{n'l'}}{\pd t} -u_{n'l'}^*\frac{\pd u_{nl}}{\pd t}}
        \end{split}
    \end{equation}
    For $n=n',l=l'$, this leads to
    \begin{equation}
        2\abs{\bk_{nl}}|N_{nl}|^2\int_0^L\!\!\dd x\int_0^L\!\!\dd y\sin^2\frac{n\pi y}{L}= \abs{\bk_{nl}}|N_{nl}|^2L^2 = 1
    \end{equation}
    and hence we can set $N_{nl} = 1/\sqrt{2L\abs{\bk_{nl}}}$.
    
\begin{figure}
        \centering
        \includegraphics[scale=1]{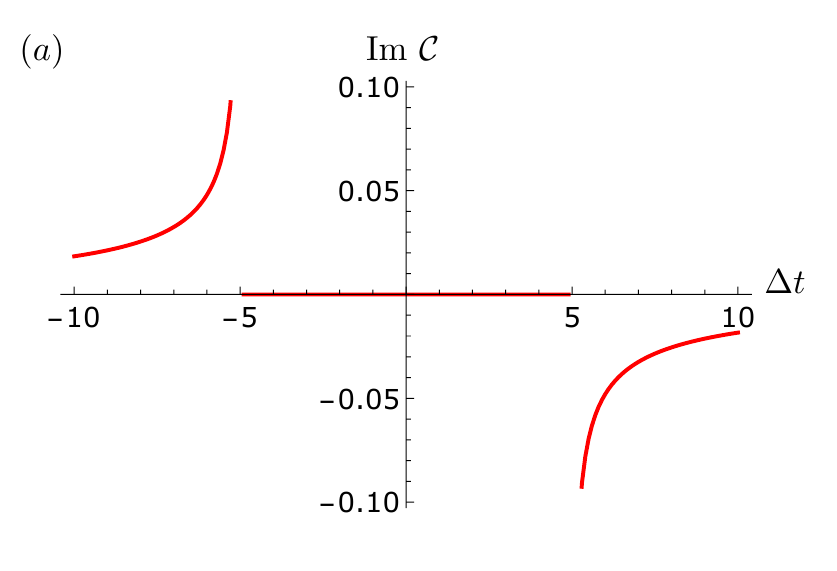}
        \includegraphics[scale=1]{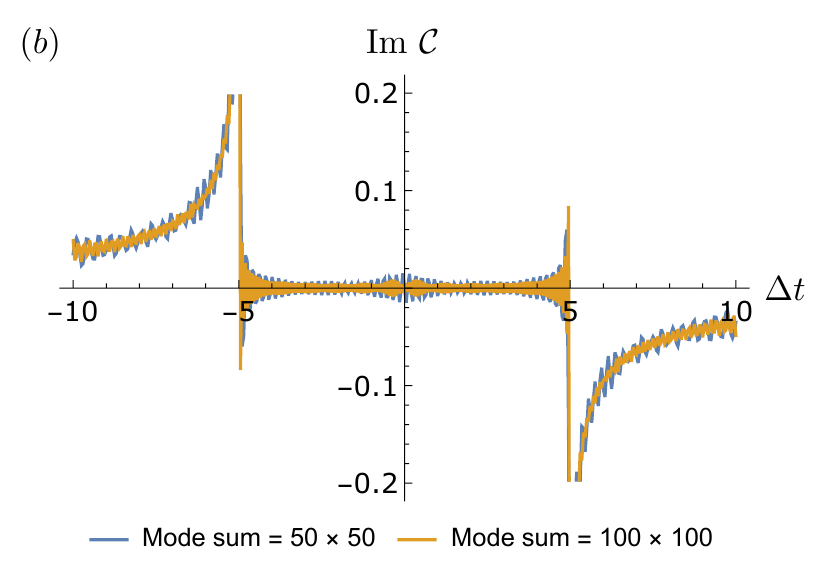}
        \caption{\textbf{(a)} Free space commutator. \textbf{(b)} Commutator for finite cylindrical spacetime with spatial topology $\Sigma=I\times S^1$ for $50\times 50$ and $100\times 100$ modes. Within the timelike interval, as we sum more higher mode, the estimator uniformly approaches zero (here the average is already zero). At the null boundary there is Gibbs phenomenon due to the UV cutoff.} 
        \label{fig:2Dcylinder}
    \end{figure}

    The above expression alone is sufficient to conclude that there is no zero mode problem even though we have $n=0$ eigensolutions. The reason is because since $l\in \mathbb{N}$, we have $\abs{\bk_{nl}}\neq 0$ for all $n\in \mathbb{Z}$ including $n=0$. Consequently, under canonical quantization every mode with a definite $n,l$ is an oscillator mode with nonzero frequency $\abs{\bk_{nl}}$. Without computing the commutator, we will know that microcausality is fully governed by the oscillator modes. We show this concretely in Figure~\ref{fig:2Dcylinder}, where we highlight the differences between the signalling of the detectors in free space studied in \cite{Causality2015Eduardo} (Fig.~4a) and detectors in finite cylindrical cavity of topology $I\times S^1$ (Fig.~4b). 
    
    To have a zero mode in $(2+1)$ dimensions, we must consider the case where we have the ``harmonic" solution with vanishing frequency $\abs{\bk_{00}}=0$. This suggests two other nontrivial cases: (1) toroidal boundary and (2) a $(2+1)$ dimensional Einstein cylinder. 
    
    \subsection{Toroidal boundary condition: $\Sigma=S^1\times S^1$}
    For the case with toroidal boundary, the spatial topology is $S^1\times S^1$, i.e.  both $x,y$ have periodic boundary conditions,
    \begin{equation}
        \begin{split}
            \hat\phi(t,x,y) &= \hat\phi(t,x+L_1,y)\,,\\
            \hat\phi(t,x,y) &= \hat\phi(t,x,y+L_2)\,.
        \end{split}
    \end{equation}
    Again for simplicity let us set $L_1=L_2=L$. This gives us the positive frequency eigenmodes of the form
    \begin{equation}
    \begin{split}
        u_{mn}(t,x,y) &= N_{mn}e^{-\ii\abs{\bk_{mn}}t}\exp\frac{2\ii\pi m  x}{L}\exp\frac{2\ii\pi n  y}{L}\,,\\
        \abs{\bk_{mn}} &= \sqrt{\rr{\frac{2\pi m}{L}}^2+\rr{\frac{2\pi n}{L}}^2}\,,\\
        N_{mn} &= \frac{1}{\sqrt{2L\abs{\bk_{mn}}}}\,,
    \end{split}
    \end{equation}
   where in this case the zero mode will appear. The oscillator part of the commutator is given by (see Appendix~\ref{appendix: commutators})
    \begin{equation}
        \begin{split}
        &\braket{[\hat\phi_{\text{osc}}(\sx),\hat\phi_{\text{osc}}(\sx')]}\\
        &= \sum_{m=-\infty}^\infty\sum_{n\neq 0} \frac{1}{\sqrt{2L\abs{\bk_{mn}}}}\rr{u_{mn}{u'}^*_{mn}-u
        '_{mn}u^*_{mn}} \, +\\
        &\hspace{0.6cm}\sum_{m\neq 0}     \frac{1}{\sqrt{2L\abs{\bk_{m0}}}}\rr{u_{m0}{u'}^*_{m0}-u
       '_{m0}u^*_{m0}}\,.
        \end{split}
    \end{equation}
    Since the sum cannot be done analytically, we resort to partial sums for the computation of estimator $\mathcal{C}$ and take the imaginary part, $\text{Im } \mathcal{C}$. This will give us the same information about superluminal signalling due to the absence of zero mode, since from Eq.~\eqref{eq:estimator} and for a delta-switching and pointlike detector the estimator $\mathcal{E}$ is the modulus of $\mathcal{C}$, which is purely imaginary. Plotting  $\text{Im } \mathcal{C}$ is visually clearer.
    
\begin{figure}
        \centering
        \includegraphics[scale=1]{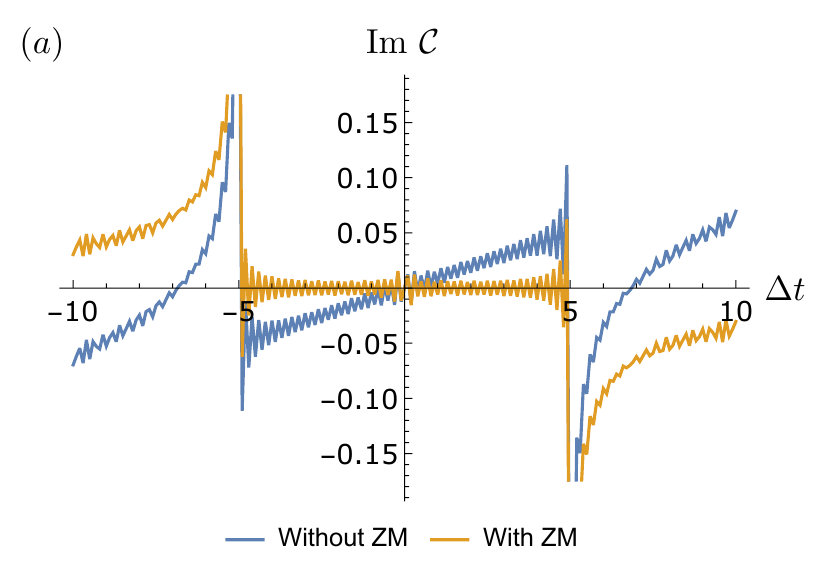}
        \includegraphics[scale=1]{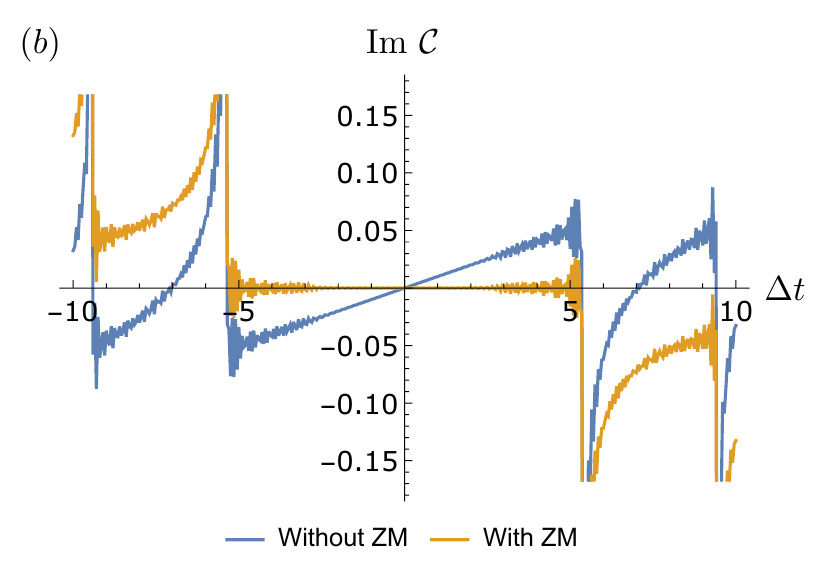}
        \caption{Commutator for toroidal spacetime with spatial topology $\Sigma=S^1\times S^1$. \textbf{(a)}  $50\times 50$ modes, $\Delta x = 5,\Delta y = 0$. \textbf{(b)} for $100\times 100$ modes, $\Delta x = 5,\Delta y = 2 $.}
        \label{fig:2Dtorus}
    \end{figure}

    The results are shown in Figure~\ref{fig:2Dtorus}-a). It is clear that there is a causality violation and superluminal signalling between detectors when one discounts zero mode. Causality is recovered  when the zero mode contribution is included, even at the level of partial sums. Furthermore, note that the zero mode commutator is different from the one in $(1+1)$ dimensions, namely
    \begin{equation}
        [\hat\phi_\text{zm}(\sx),\hat\phi_\text{zm}(\sx')] = -\frac{\ii\Delta t}{L^2}\,.
    \end{equation}

    \subsection{(2+1) dimensional Einstein cylinder}
    
\begin{figure}[tp]
        \includegraphics[scale=1]{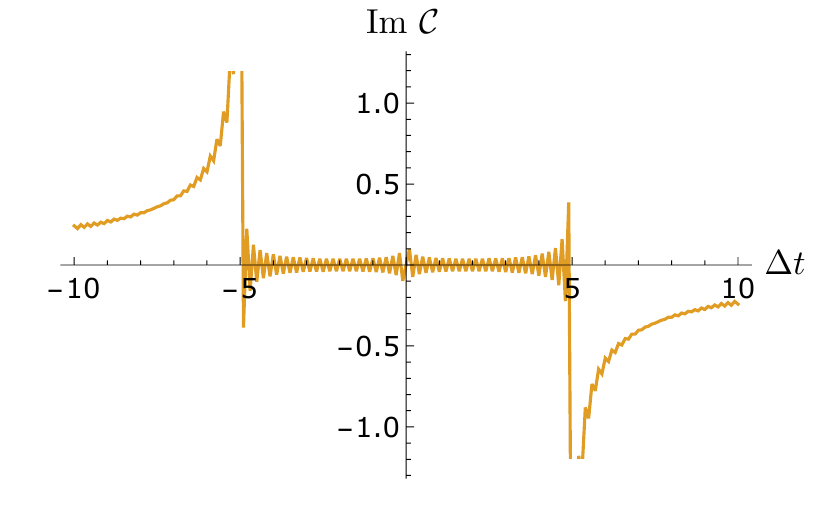}
        \caption{Commutator for $(2+1)$ dimensional Einstein cylindrical spacetime with spatial topology $\Sigma=\R \times S^1$ for $(k_{\min},k_{\max}) = (-50,50)$. It does not display causality violation despite the integral domain excluding the zero mode.}
        \label{fig:2DEinstein}
    \end{figure}

    The other nontrivial case involves the  Einstein cylinder, where the only difference is that the sum over modes along one direction is a continuum (hence an integral over modes instead of a summation). The mode decomposition is given by
    \begin{equation}
    \begin{split}
        u_{nl}(t,x,y) &= N_{nl}e^{-\ii\abs{\bk_{nl}}t}e^{\ii ly}\exp\frac{2\ii\pi n x}{L}\,,\\
        \abs{\bk_{nl}} &= \sqrt{\rr{\frac{2\pi n}{L}}^2+l^2}\,,\hspace{0.5cm}n\in\mathbb{Z},l\in \mathbb{R}\,,\\
        N_{nl} &= \frac{1}{\sqrt{2L\abs{\bk_{nl}}}}\,.
    \end{split}
    \end{equation}
    While formally it appears that the result should be the same as the case for toroidal scenario, we should be careful because from the perspective of the $y$-modes, $\omega_{00}$-mode is a point and hence is a measure zero proper subset of the real line $\R$ which has strictly greater measure.
    The partially mode-summed estimator is shown in Figure~\ref{fig:2DEinstein}, where we can see the faster-than-light signalling that appears when the zero mode is ignored. Indeed one can check that taking the principal value integral for the full commutator 
    \begin{equation}
    \begin{split}
    \label{eq:comm2Deinstein}
        &[\hat \phi(\sx),\hat\phi(\sx')] \\
        &= \sum_{n\in \mathbb{Z}}\int_{-\infty}^{-\epsilon}\!\!\! \dd l \left(u_{nl}(\sx)u^*_{nl}(\sx')-u_{nl}(\sx')u^*_{nl}(\sx)\right) + \\
        &\hspace{0.4cm}\sum_{n\in \mathbb{Z}}\int_\epsilon^{\infty}\!\!\!  \dd l \left(u_{nl}(\sx)u^*_{nl}(\sx')-u_{nl}(\sx')u^*_{nl}(\sx)\right)
    \end{split}
    \end{equation}
    does not violate microcausality as $\epsilon\to 0$.
    
    Again we note that Neumann boundary conditions similarly produce a zero mode, as in $(1+1)$ dimensions. However, in $(2+1)$ dimensions it is now possible to have periodic boundary conditions in one direction and Neumann boundary conditions on another. A zero mode will arise whenever there is ``zero frequency" component of the eigenfunctions which is spatially constant (see Appendix~\ref{appendix: commutators} and Appendix~\ref{appendix: zerocommutator} for more details).
    
    \section{Results in higher dimensions}
    \label{sec:discuss} 
    
\begin{figure}[tp]
        \includegraphics[scale=1]{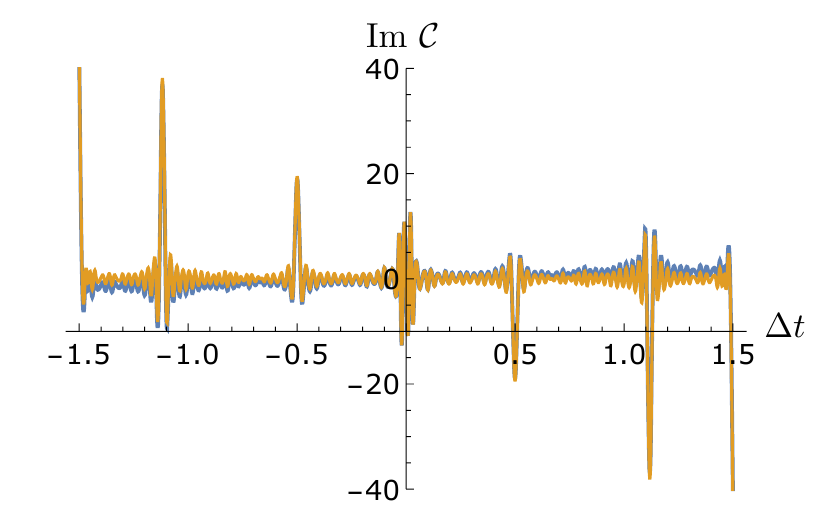}
        \caption{Commutator for $(3+1)$ dimensional toroidal spacetime with spatial topology $\Sigma=S^1\times S^1 \times S^1$ for $30\times30\times30$ oscillator modes. Here $L=1$ and $\Delta x = 0.5,\Delta y=\Delta z = 0$. The curve without zero mode is somewhat tilted clockwise relative to the origin, reflecting causality violation. The spikes correspond to the divergences due to the support of the commutator on the null cone.}
        \label{fig:3Dtorus}
    \end{figure}

    Based on our results in $(2+1)$ dimensions, we can easily generalize the results to higher dimensions. In particular, the toroidal case with topology $S^1\times S^1\times ... \times S^1$ will present a zero mode in arbitrary dimensions since the construction is analogous. The oscillator part of the commutator for arbitrary dimensions with toroidal boundary conditions (and more general boundary conditions) is given by Eq.~\eqref{eq: generalcommutator} in Appendix~\ref{appendix: commutators}. Another notable feature is that in higher dimensions, one can have strong Huygens' principle, e.g. in $(3+1)$ dimensions \cite{Sonego1992huygens,Valerio1999tails}. This is shown in Figure~\ref{fig:3Dtorus}, where the support of the full commutator (including the zero mode) is only on the light cone. Notice that the zero mode commutator in arbitrary dimensions is given by (see Appendix~\ref{appendix: zerocommutator} for derivation)
    \begin{equation}
        [\hat\phi_{\text{zm}}(\sx),\hat\phi_{\text{zm}}(\sx')] = -\frac{\ii\Delta t}{L^n}\,, \hspace{0.5cm}\Delta t = t-t'\,.
    \end{equation}
    That is, the impact of the zero mode is polynomially weaker in higher dimensions. In Figure~\ref{fig:3Dtorus} we already see that the estimator $\text{Im }\mathcal{C}$ is not very much different visually, but removing zero mode nonetheless leads to causality violation  and, in this case, also violation of strong Huygens' principle  in $(3+1)$ dimensions.
    
    Another feature of higher dimensional cases is that there are more transverse dimensions in which one can impose boundary conditions. For example, to have zero mode, strictly speaking one does not need toroidal boundary condition. One could instead use a combination of periodic boundary condition in some transverse dimensions and Neumann boundary condition on the remaining dimensions (see Appendix~\ref{appendix: commutators} for details).

    \section{Conclusion}
    
    In this paper we have shown that what a zero mode is present due to periodic or Neumann boundary conditions (associated either to cavities or spacetimes with compact spatial topology) excluding them in modelling light-matter interactions using particle detector models can lead to faster-than-light signalling between two detectors. 
    
    We explicitly quantify the amount of violation in terms of the strength of the superluminal signal that one emitter operating a particle detector can send to another if the detector is not coupling to the zero mode, and find that for a fixed spatial separation $\Delta \bx$, the causality violation decays polynomially with the temporal separation $\Delta t$, and the length across the boundary condition $L$. The power law of this decay is given by the number of spatial dimensions $n$ so that the decay is linear in $(1+1)$ dimensions, quadratic in $(2+1)$, etc. Therefore, any relativistic scenario where we analyze the light-matter interaction, communication, entanglement harvesting, or any other phenomenological study where relativity is of importance should consider that particle detectors couple to the zero mode explicitly. 
    As a corollary, in such scenarios one might need to care about the state of the zero mode, whose impact on detector dynamics is non-trivial \cite{Zeromode2014Edu}, and one cannot get around the ambiguity of establishing the state of a zero mode just by ignoring its presence.

    \section*{Acknowledgment}
    The authors thank Jorma Louko for useful discussions during the Relativistic Quantum Information-North 2018 conference in Vienna. E.T. acknowledges support of Mike-Ophelia Lazaridis Fellowship.  E.M.-M. acknowledges funding of the Natural Sciences and Engineering Research Council of Canada (NSERC) Discovery program and his Ontario Early Researcher grant.

    \appendix
    
    \section{Derivation of the oscillator part of the commutator $\braket{[\hat\phi_{\text{osc}}(\sx),\hat\phi_{\text{osc}}(\sx')]}$}
    \label{appendix: commutators}
    In this Appendix we will write derive the expressions for the commutators of the field in arbitrary dimensions when periodic boundary conditions and Neumann boundary conditions are imposed, or a combination of periodic and Neumann boundary conditions if spatial dimension is at least two. We first derive the most general expression and then illustrate in full detail the particular cases for various boundary conditions in $(1+1)$ and $(2+1)$ dimensions.
    
    \subsection{General expression in arbitrary dimensions}
    
  `In  $(n+1)$ dimensions, given an arbitrary state of the field $\hat\rho_{\hat\phi}$, the expectation value of the commutator with respect to the state $\field$ is given by
    \begin{equation}
        \braket{[\hat\phi_{\text{osc}}(\sx),\hat\phi_{\text{osc}}(\sx')]}_{\hat \rho} = \tr\left(\field[\hat\phi_{\text{osc}}(\sx),\hat\phi_{\text{osc}}(\sx')]\right)\,.
    \end{equation}
    To reduce notational clutter, let us define $I,J$ to be collective indices where $n$ is the number of spatial dimensions. This will simplify the expression for the sum over modes in the commutators below. We define $\allA:=A_1,A_2,...,A_n$ to be the collective indexing sets for $I,J$ which excludes the zero mode (if any). That is, $I\in \allA$ means that every  component $i_l$ of the multi-index $I$ takes values in the set $A_l$ for each $l=1,2,...,n$.

    We can expand the field operator in terms of a complete set of orthonormal solutions to the Klein-Gordon equation $\{u_{I},u^*_{I}\}$, that is,
    \begin{align}
        \hat\phi(\sx) = \sum_{I\in \allA}\left(\hat a_I u_I(\sx)+\hat a_I^\dagger u_I^*(\sx)\right)\,.
    \end{align}
    Notice that the sum over the set of modes $I$ can be a continuous sum (an integral) or a discrete sum depending on the boundary conditions imposed (by changing the indexing set $\allA$). We can now evaluate then the expectation of the commutator as 
    \begin{widetext}
        \begin{align}
            \braket{[\hat\phi_{\text{osc}}(\sx),\hat\phi_{\text{osc}}(\sx')]}_{\hat{\rho}} \notag &= \sum_{I,J\in\allA}\left(\braket{[\hat a_I,\hat a_{J}]}_{\hat{\rho}}u_Iu_J' + \braket{[\hat a_I^\dagger,\hat a_J^\dagger]}_{\hat{\rho}}u_I^*{u'_J}^{*} \, +\braket{[\hat a_I,\hat a_J^\dagger]}_{\hat{\rho}}u_I{u'_J}^{*} + \braket{[\hat a_I^\dagger,\hat a_J]}_{\hat{\rho}}u_I^*u_J'\right)\notag\\
            &= \sum_{I,J\in\allA}\left( \delta_{IJ} u_I{u'_J}^{*} - \delta_{JI} u_I^*u_J'\right)\notag\\
            &= \sum_{I\in\allA}\left( u_I(\sx)u_I^{*}(\sx') - u_I^*(\sx)u_I(\sx')\right)
            \notag\\
            &\equiv\braket{[\hat\phi_{\text{osc}}(\sx),\hat\phi_{\text{osc}}(\sx')]} \label{eq: generalcommutator}\,,
        \end{align}
    \end{widetext}
    where we have shortened notation by using $u_I'\equiv u_I(\sx')$. We have also used the canonical commutation relations $[\hat a_I,\hat a_J^\dagger]=\delta_{IJ}\openone$ to show explicitly the fact that the expectation value of the commutator is independent of the state of the field and drop the subscript $\hat \rho$ from the expectation value. 
    
    Eq.~\eqref{eq: generalcommutator} above is the most general expression for the commutator of the oscillator part of the field. When different boundary conditions are imposed, we vary the choice of indexing set $\allA$. For example, in the case of toroidal boundary conditions (periodic in all $n$ spatial dimensions), the eigenfunctions are given by
    \begin{equation}
        u_I = N_I e^{-\ii |\bk_I|t+\ii\bk_I\cdot \bx}\,,\abs{\bk_I} = \sqrt{\sum_{l=1}^n\rr{\frac{2\pi i_l}{L}}^2}\,,
    \end{equation}
    where the indices $i_l$ are nonzero integers (hence excludes the zero mode), i.e., $A_l=\mathbb{Z}\setminus\{0\}$. That is, the oscillator part of the commutator reads
    \begin{widetext}
    \begin{align}
           \braket{ [\hat\phi_{\text{osc}}(\sx),\hat\phi_{\text{osc}}(\sx')]}
            &=  \sum_{j_1\in \mathbb{Z}}...\sum_{j_{n-1}\in \mathbb{Z}}\sum_{j_n\neq 0} u_{j_1...j_n}(\sx)u_{j_1...j_n}^*(\sx') - u_{j_1...j_n}(\sx')u_{j_1...j_n}^*(\sx)\, + \notag \\
            &\hspace{0.4cm}\sum_{j_1\in \mathbb{Z}}...\sum_{j_{n-2}\in \mathbb{Z}}\sum_{j_{n-1}\neq 0} u_{j_1...j_{n-1}0}(\sx)u_{j_1...j_{n-1}0}^*(\sx') - u_{j_1...j_{n-1}0}(\sx')u_{j_1...j_{n-1}0}^*(\sx) \,+ \notag\\
            &\hspace{0.4cm}\sum_{j_1\in \mathbb{Z}}...\sum_{j_{n-3}\in \mathbb{Z}}\sum_{j_{n-2}\neq 0} u_{j_1...j_{n-2}00}(\sx)u_{j_1...j_{n-2}00}^*(\sx') - u_{j_1...j_{n-2}00}(\sx')u_{j_1...j_{n-2}00}^*(\sx) \,+\notag \\
            &\hspace{0.4cm}\sum_{j_1\neq 0} u_{j_100...0}(\sx)u_{j_100...0}^*(\sx') - u_{j_100...0}(\sx')u_{j_100...0}^*(\sx)\,.
    \end{align}
    \end{widetext}
    For Neumann boundary conditions, the eigenfunctions are instead given by
    \begin{align}
        u_I &= N_I \prod_{l=1}^n\cos\frac{i_l\pi x_l}{L} \frac{}{}e^{-\ii |\bk_I|t}\,,
        \abs{\bk_I} = \sqrt{\sum_{l=1}^n\rr{\frac{i_l\pi}{L}}^2}\,,
    \end{align}
    and the indexing set is given by $A_l = \mathbb{N}\cup\{0\}$ in such a way that it excludes the zero mode, i.e. at least one of the summation is over $\mathbb{N}$. More concretely, we replace the summation for $j_k\in \mathbb{Z}$ with $j_k\in \mathbb{N}\cup\{0\}$ and the summation for $j_k\neq 0$ with $j_k\in \mathbb{N}$ in \eqref{eq: generalcommutator}.
    
    We will now use these results to write down the explicit expressions used in this paper.
    
    \begin{widetext}
    \subsection{(1+1)  periodic boundary conditions}
    
    For periodic boundary conditions, the eigenfunctions of the Klein-Gordon equation read
    \begin{equation}
        \begin{split}
        u_n(t,x) &= \frac{1}{\sqrt{2\pi n}}e^{-\ii|k_n|t+\ii k_nx}\,, k_n = \frac{2\pi n}{L}\,, n\neq 0\,.
    \end{split}
    \end{equation}
    where the normalization constant $N_n=1/\sqrt{2\pi n}$. The expectation of the commutator reads
    \begin{equation}
    \begin{split}
       \braket{[\hat\phi_{\text{osc}}(\sx)\hat\phi_{\text{osc}}(\sx')]} &= \sum_{n=1}^\infty \frac{1}{4\pi n}\left[e^{-\ii k_n(\Delta u-\ii\epsilon)}+e^{-\ii k_n(\Delta v-\ii\epsilon)}\right.-\left.e^{\ii k_n(\Delta u-\ii\epsilon)}-e^{\ii k_n(\Delta v-\ii\epsilon)}\right]\\
       &= \sum_{n=1}^\infty \frac{\ii}{2\pi n}\left[\sin\frac{2\pi n}{L}(\Delta u-\ii\epsilon)+\sin\frac{2\pi n}{L}(\Delta v-\ii\epsilon)\right]\,, 
      \end{split}
    \end{equation}
    where $u=t-x$ and $v=t+x$ are the double null coordinates. Finally, we invoke the following identity
    \begin{equation}
        \sum_{n=1}^\infty \frac{\sin nx}{n} = \frac{1}{2} \left[\log \left(1-e^{-\ii x}\right)-\log \left(1-e^{\ii x}\right)\right]
    \end{equation}
    and we obtain the commutator in Eq.~\eqref{eq:osccommutator1D}. 
    
    \subsection{(1+1)  Neumann boundary conditions}
    
    In the case of Neumann boundary condition, the eigenfunctions of the Klein-Gordon equation take the form
    \begin{equation}
    \begin{split}
    \label{eq:neumann1Deigen}
        u_n(t,x) &= \frac{1}{\sqrt{n\pi}}\cos\frac{n\pi x}{L}e^{-\ii|k_n|t}\,, k_n = \frac{n\pi}{L}\,, n\in \mathbb{N}\,,
    \end{split}
    \end{equation}
    where the normalization constant $N_n=1/\sqrt{\pi n}$. The expectation of the commutator reads
    \begin{align}
       \braket{[\hat\phi_{\text{osc}}(\sx)\hat\phi_{\text{osc}}(\sx')]}\notag &= \sum_{n=1}^\infty \frac{1}{\pi n}\left[\cos{\frac{n\pi x}{L}}\cos{\frac{n\pi x'}{L}}e^{-\ii k_n(\Delta t-\ii\epsilon)}\right. -\left.\cos{\frac{n\pi x'}{L}}\cos{\frac{n\pi x}{L}}e^{\ii k_n(\Delta t-\ii\epsilon)}\right]\notag\\
       &= \sum_{n=1}^\infty \frac{-2i}{\pi n}\left[\cos{\frac{n\pi x}{L}}\cos{\frac{n\pi x'}{L}}\sin\frac{n\pi \Delta t}{L}\right]\,,
    \end{align}
    where $\Delta t = t-t'$. Notice that due to the form of the eigenfunction in Eq.~\eqref{eq:neumann1Deigen}, this commutator is no longer translation-invariant, unlike the case of periodic boundary conditions where the mode sums are purely functions of $\Delta u$ and $\Delta v$. Still, this expression admits a closed analytic expression, namely:
    \begin{widetext}
    \begin{equation}
        \begin{split}
         &\braket{[\hat\phi_{\text{osc}}(\sx)\hat\phi_{\text{osc}}(\sx')]}\\
         &=\frac{1}{4\pi}\left[
         \log \left(1-e^{\frac{\ii \pi(u-v'-\ii\epsilon)}{L}}\right)+
         \log \left(1-e^{\frac{\ii \pi(\Delta v-\ii\epsilon)}{L}}\right)+
         \log \left(1-e^{\frac{\ii \pi(\Delta u-\ii\epsilon)}{L}}\right)+
         \log \left(1-e^{\frac{\ii \pi(v-u'-\ii\epsilon)}{L}}\right)\right.\\
         &\hspace{0.75cm}\left.
         -\log \left(1-e^{-\frac{i \pi (u-v'-\ii\epsilon)}{L}}\right)
         -\log \left(1-e^{-\frac{i \pi (\Delta v-\ii\epsilon)}{L}}\right)
         -\log \left(1-e^{-\frac{i \pi (\Delta u-\ii\epsilon)}{L}}\right)
         -\log \left(1-e^{-\frac{i \pi (v-u'-\ii\epsilon)}{L}}\right)
         \right].
        \end{split}
    \end{equation}
    \end{widetext}

    \subsection{(2+1) dimensions periodic boundary conditions}
    
    In $(n+1)$ dimensions with $n\geq 2$, the mode sums do not have a closed form because the normalization constant $N_I$ mixes contributions from different transverse momenta. As such, in practice, one would numerically impose a UV cutoff to evaluate these sums.
    
    For simplicity let us impose the boundary condition across a length $L$ in both transverse directions. This will simplify the expression for the normalization constant $N_{mn}$. The eigenfunctions for toroidal boundary condition (periodic boundary in both spatial directions) in $(2+1)$ dimensions read
    \begin{align}
        u_{mn}(t,x,y)&= N_{mn}\exp\left[-\ii|\bm{k}_{mn}|t+\ii\frac{2\pi m}{L}x+\ii\frac{2\pi n}{L}y\right]\,,\notag\\
        \bm{k}_{mn} &= \sqrt{\rr{\frac{2\pi m}{L}}^2+\rr{\frac{2\pi n}{L}}^2}\,.
    \end{align}
    Recall that the normalization constant $N_{mn}$ couples momenta from both transverse directions,
    \begin{equation}
        N_{mn} = \frac{1}{\sqrt{L\abs{\bk_{mn}}}}\,.
    \end{equation}
    The expectation value of the commutator $\braket{[\hat\phi_{\text{osc}}(\sx)\hat\phi_{\text{osc}}(\sx')]}$ is then given by the following sum:
    \begin{align}
       &\braket{[\hat\phi_{\text{osc}}(\sx)\hat\phi_{\text{osc}}(\sx')]}\notag\\
       &= \sum_{m=-\infty}^\infty\sum_{n\neq 0} \frac{1}{\sqrt{2L\abs{\bk_{mn}}}}\rr{u_{mn}{u'}^*_{mn}-u
       '_{mn}u^*_{mn}} \, +\notag\\
       &\hspace{0.6cm}\sum_{m\neq 0} \frac{1}{\sqrt{2L\abs{\bk_{m0}}}}\rr{u_{m0}{u'}^*_{m0}-u
       '_{m0}u^*_{m0}}\,.
    \end{align}
    These two sums only exclude the $\abs{\bk_{00}}$ term corresponding to the zero mode. This expression generalizes easily to higher dimensions, essentially including all sums which excludes the ``zero frequency" part containing $\abs{\bk_{00...0}}$. 
    
    \end{widetext}
    
    \subsection{(2+1) dimensions Neumann boundary conditions}
    
    For the Neumann boundary condition on both transverse directions, we get
    \begin{equation}
        \begin{split}
            u_{mn}(t,x,y)&= N_{mn}\cos\frac{m\pi x}{L}\cos\frac{n\pi y}{L}e^{-\ii|\bm{k}_{mn}|t}\,,\\
        \bm{k}_{mn} &= \sqrt{\rr{\frac{\pi m}{L}}^2+\rr{\frac{\pi n}{L}}^2}\,.
        \end{split}
    \end{equation}
    According to the prescription in Eq.~\eqref{eq: generalcommutator}, the expectation of the commutator will now read
    \begin{align}
       &\braket{[\hat\phi_{\text{osc}}(\sx)\hat\phi_{\text{osc}}(\sx')]}\notag\\
       &= \sum_{m=0}^\infty\sum_{n=1}^\infty \frac{1}{\sqrt{2L\abs{\bk_{mn}}}}\rr{u_{mn}{u'}^*_{mn}-u
       '_{mn}u^*_{mn}} \, +\notag\\
       &\hspace{0.4cm}\sum_{m=1}^\infty \frac{1}{\sqrt{2L\abs{\bk_{m0}}}}\rr{u_{m0}{u'}^*_{m0}-u
       '_{m0}u^*_{m0}}\,.
    \end{align}
    In fact, this suggests the possibility of using periodic \textit{and} Neumann boundary conditions on different transverse dimensions. If we impose Neumann boundary along $x$-direction and periodic boundary across $y$-direction, the eigenfunctions would be
    \begin{equation}
        \begin{split}
            u_{mn}(t,x,y)&= N_{mn}\cos\frac{m\pi x}{L}\exp\left[-\ii|\bm{k}_{mn}|t + \ii\frac{2\pi ny}{L}\right]\,,\\
        \bm{k}_{mn} &= \sqrt{\rr{\frac{\pi m}{L}}^2+\rr{\frac{2\pi n}{L}}^2}\,,
        \end{split}
    \end{equation}
    where $m\in \mathbb{N}\cup \{0\}$ and $n\in \mathbb{Z}$. The expected value of the commutator will now take the form
    \begin{align}
       &\braket{[\hat\phi_{\text{osc}}(\sx)\hat\phi_{\text{osc}}(\sx')]}\\
       &= \sum_{m=0}^\infty\sum_{n\neq 0} \frac{1}{\sqrt{2L\abs{\bk_{mn}}}}\rr{u_{mn}{u'}^*_{mn}-u
       '_{mn}u^*_{mn}} \, +\\
       &\hspace{0.4cm}\sum_{m=1}^\infty \frac{1}{\sqrt{2L\abs{\bk_{m0}}}}\rr{u_{m0}{u'}^*_{m0}-u
       '_{m0}u^*_{m0}}\,.
    \end{align}

    \section{(\textit{n}+1) Einstein cylinder}

    In the case of $(n+1)$ Einstein cylinder, the result is analogous to the toroidal case except replacing the sum over $\mathbb{Z}$ with an integral over momentum along the non-compact spatial direction (see Eq.~\eqref{eq:comm2Deinstein} for the $(2+1)$ case). In our multi-index notation, this is basically setting $A_l=\mathbb{R}$ for non-compact transverse dimensions and integrating over momentum instead of summing over discrete momentum. However, since the spectrum is continuous, the commutator of the oscillator modes computed in this manner is in fact the full field commutator (or rather, the zero mode does not contribute since it is a point of measure zero in momentum space). Therefore, effectively there is no zero mode relevant physics in the Einstein cylinder when $n\geq 2$.

    \section{Derivation of the zero mode commutator $\braket{[\hat\phi_{\text{zm}}(\sx),\hat\phi_{\text{zm}}(\sx')]}$}

    \label{appendix: zerocommutator}
    Here we derive the fact that the zero mode commutator scales polynomially with the length of the ``cavity" where the boundary conditions are imposed, i.e.
    \begin{equation}
        \braket{[\hat\phi_{\text{zm}}(\sx),\hat\phi_{\text{zm}}(\sx')]} = -\ii\frac{\Delta t}{L^n}
    \end{equation}
    where $n$ is the number of spatial dimensions. Thus, in some sense, the zero mode contribution is (polynomially) weaker in higher dimensions.
    
    To prove this, it is simplest to start from the Lagrangian of the field theory. In $(n+1)$ dimensions, the Lagrangian is given by
    \begin{equation}
        \begin{split}
        \label{eq:lagrangian}
        \mathcal{L} &= \frac{1}{2}\int\dd^n\bx \,\pd_\mu\phi(t,\bx)\pd^\mu\phi(t,\bx)\\
        &= \frac{1}{2}\int\dd^n\bx \,\left[\rr{\frac{\pd\phi}{\pd t}}^2+ \rr{\nabla\phi}^2\right]\,.
        \end{split}
    \end{equation}
    The boundary conditions which will produce zero modes need to have discrete spectrum. Hence, the field can be expanded as a Fourier series
    \begin{equation}
        \phi(t,\bx) = \sum_{I\in\allA} \varphi_{_I}(t) e^{\ii\bk_I\cdot \bx}
    \end{equation}
    where we have used the notation $I$ for collective indices for summation as defined in Appendix~\ref{appendix: commutators}. Here we denote the Fourier coefficients as $\{\varphi_I(t)\}$.
    
    The first in term in Eq.~\eqref{eq:lagrangian} reads
    \begin{equation}
        \begin{split}
        \int\dd^n\bx \rr{\frac{\pd\phi}{\pd t}}^2 &= \int\dd^n\bx  \sum_{I\in\allA}\sum_{J\in\allA}\dot{\varphi}_{_I}(t)\dot{\varphi}_{_J}(t)e^{\ii(\bk_I+\bk_J)\cdot\bx}\,.
        \end{split}
    \end{equation}
    When we have periodic/Neumann boundary conditions across distance $L$ (in all spatial dimensions), the expression becomes
    \begin{align}
        \int\dd^n\bx \rr{\frac{\pd\phi}{\pd t}}^2 
        &= \int_{[0,L]^n}\!\!\!\!\!\!\!\!\!\!\dd^n\bx \sum_{I,J\in\allA}\dot{\varphi}_{_I}\dot{\varphi}_{_J}e^{\ii(\bk_I+\bk_J)\cdot\bx}\notag\\
        &= L^n\sum_{I\in\allA}\dot{\varphi}_{_I}\dot{\varphi}_{_{-I}}\,.
    \end{align}
    The second term reads
    \begin{equation}
        \begin{split}
        \int_{[0,L]^n}\!\!\!\!\!\!\!\!\!\!\dd^n\bx \rr{\nabla\phi}^2 
        &= -|\bk_I|^2L^n\sum_{I\in\allA} \varphi_{_I}\varphi_{_{-I}}\,.
        \end{split}
    \end{equation}
    The full Lagrangian is therefore given by
    \begin{equation}
        \begin{split}
        \mathcal{L} &= \frac{L^n}{2}\sum_{I\in\allA}\left[\dot{\varphi}_{_I}\dot{\varphi}_{_{-I}}-|\bk_I|^2 \varphi_{_I}\varphi_{_{-I}}\right]\,.
        \end{split}
    \end{equation}
    From this expression, we can read off the zero mode Lagrangian (which corresponds to $|\bk_I| = 0$ with $I = j_1j_2...j_n =  00...0$), namely
    \begin{equation}
    \begin{split}
        \mathcal{L}_{\text{zm}} &= \frac{L^n}{2}\dot\varphi_{0...0}^2 \equiv \frac{L^n}{2}\dot{Q}^2\,.
    \end{split}
    \end{equation}
    The case for $n=1$ is given in \cite{francesco1996conformal,Zeromode2014Edu}. The momentum conjugate to $\varphi_I$ is given by
    \begin{equation}
        \pi_{_{I}} = \frac{\pd \mathcal{L}}{\pd(\dot{\varphi_{_I}})} = L^n\dot\varphi_{_{-I}}\,,
    \end{equation}
    hence the Hamiltonian is given by
    \begin{equation}
        \begin{split}
        H &= \left[\sum_{I\in\allA} \frac{ \pi_{_I} \pi_{_{-I}}}{2L^n}  -\frac{|\bk_I|^2}{2}\sum_{I\in\allA} \varphi_{_I}\varphi_{_{-I}}\right]\,.
        \end{split}
    \end{equation}
    Canonical quantization converts $\pi_I$, $\varphi_I$ into operators $\hat \pi_I,\hat\varphi_I$, thus we have the zero mode Hamiltonian in $(n+1)$ dimensions:
    \begin{equation}
        \begin{split}
            \hat H_{\text{zm}} &= \frac{\hat P_{0...0}^2}{2L^n} \equiv \frac{\hat P^2}{2L^n}\,.
        \end{split}
    \end{equation}
   So, for $n$ dimensions, the procedure that lead Eq.\eqref{eq:1Dzero} is exactly the same replacing $L$ by $L^n$ in~\eqref{eq:1Dzerohamiltonian}. Consequently, the commutator of the zero mode in $(n+1)$ dimension is obtained by replacing $L$ with $L^n$, namely
    \begin{equation}
        \braket{[\hat\phi_{\text{zm}}(\sx),\hat\phi_{\text{zm}}(\sx')]} = -\frac{\ii\Delta t}{L^n}\,, \hspace{0.5cm}\Delta t = t-t'\,,
    \end{equation}
    as claimed.

\bibliography{zeromoderef}

\begin{thebibliography}{38}%
\makeatletter
\providecommand \@ifxundefined [1]{%
 \@ifx{#1\undefined}
}%
\providecommand \@ifnum [1]{%
 \ifnum #1\expandafter \@firstoftwo
 \else \expandafter \@secondoftwo
 \fi
}%
\providecommand \@ifx [1]{%
 \ifx #1\expandafter \@firstoftwo
 \else \expandafter \@secondoftwo
 \fi
}%
\providecommand \natexlab [1]{#1}%
\providecommand \enquote  [1]{``#1''}%
\providecommand \bibnamefont  [1]{#1}%
\providecommand \bibfnamefont [1]{#1}%
\providecommand \citenamefont [1]{#1}%
\providecommand \href@noop [0]{\@secondoftwo}%
\providecommand \href [0]{\begingroup \@sanitize@url \@href}%
\providecommand \@href[1]{\@@startlink{#1}\@@href}%
\providecommand \@@href[1]{\endgroup#1\@@endlink}%
\providecommand \@sanitize@url [0]{\catcode `\\12\catcode `\$12\catcode
  `\&12\catcode `\#12\catcode `\^12\catcode `\_12\catcode `\%12\relax}%
\providecommand \@@startlink[1]{}%
\providecommand \@@endlink[0]{}%
\providecommand \url  [0]{\begingroup\@sanitize@url \@url }%
\providecommand \@url [1]{\endgroup\@href {#1}{\urlprefix }}%
\providecommand \urlprefix  [0]{URL }%
\providecommand \Eprint [0]{\href }%
\providecommand \doibase [0]{http://dx.doi.org/}%
\providecommand \selectlanguage [0]{\@gobble}%
\providecommand \bibinfo  [0]{\@secondoftwo}%
\providecommand \bibfield  [0]{\@secondoftwo}%
\providecommand \translation [1]{[#1]}%
\providecommand \BibitemOpen [0]{}%
\providecommand \bibitemStop [0]{}%
\providecommand \bibitemNoStop [0]{.\EOS\space}%
\providecommand \EOS [0]{\spacefactor3000\relax}%
\providecommand \BibitemShut  [1]{\csname bibitem#1\endcsname}%
\let\auto@bib@innerbib\@empty
\bibitem [{\citenamefont {Unruh}(1976)}]{Unruh1979evaporation}%
  \BibitemOpen
  \bibfield  {author} {\bibinfo {author} {\bibfnamefont {W.~G.}\ \bibnamefont
  {Unruh}},\ }\href {\doibase 10.1103/PhysRevD.14.870} {\bibfield  {journal}
  {\bibinfo  {journal} {Phys. Rev. D}\ }\textbf {\bibinfo {volume} {14}},\
  \bibinfo {pages} {870} (\bibinfo {year} {1976})}\BibitemShut {NoStop}%
\bibitem [{\citenamefont {Landulfo}\ and\ \citenamefont
  {Matsas}(2009)}]{Landulfo2009suddendeath}%
  \BibitemOpen
  \bibfield  {author} {\bibinfo {author} {\bibfnamefont {A.~G.~S.}\
  \bibnamefont {Landulfo}}\ and\ \bibinfo {author} {\bibfnamefont {G.~E.~A.}\
  \bibnamefont {Matsas}},\ }\href {\doibase 10.1103/PhysRevA.80.032315}
  {\bibfield  {journal} {\bibinfo  {journal} {Phys. Rev. A}\ }\textbf {\bibinfo
  {volume} {80}},\ \bibinfo {pages} {032315} (\bibinfo {year}
  {2009})}\BibitemShut {NoStop}%
\bibitem [{\citenamefont {Lopp}\ \emph {et~al.}(2018)\citenamefont {Lopp},
  \citenamefont {Mart\'in-Mart\'inez},\ and\ \citenamefont
  {Page}}]{Lopp:2018cavity}%
  \BibitemOpen
  \bibfield  {author} {\bibinfo {author} {\bibfnamefont {R.}~\bibnamefont
  {Lopp}}, \bibinfo {author} {\bibfnamefont {E.}~\bibnamefont
  {Mart\'in-Mart\'inez}}, \ and\ \bibinfo {author} {\bibfnamefont {D.~N.}\
  \bibnamefont {Page}},\ }\href
  {http://stacks.iop.org/0264-9381/35/i=22/a=224001} {\bibfield  {journal}
  {\bibinfo  {journal} {Classical and Quantum Gravity}\ }\textbf {\bibinfo
  {volume} {35}},\ \bibinfo {pages} {224001} (\bibinfo {year}
  {2018})}\BibitemShut {NoStop}%
\bibitem [{\citenamefont {Alsing}\ and\ \citenamefont
  {Fuentes}(2012)}]{Alsing2012review}%
  \BibitemOpen
  \bibfield  {author} {\bibinfo {author} {\bibfnamefont {P.~M.}\ \bibnamefont
  {Alsing}}\ and\ \bibinfo {author} {\bibfnamefont {I.}~\bibnamefont
  {Fuentes}},\ }\href {http://stacks.iop.org/0264-9381/29/i=22/a=224001}
  {\bibfield  {journal} {\bibinfo  {journal} {Classical and Quantum Gravity}\
  }\textbf {\bibinfo {volume} {29}},\ \bibinfo {pages} {224001} (\bibinfo
  {year} {2012})}\BibitemShut {NoStop}%
\bibitem [{\citenamefont {Lin}\ \emph {et~al.}(2015)\citenamefont {Lin},
  \citenamefont {Chou},\ and\ \citenamefont {Hu}}]{Lin2015teleportation}%
  \BibitemOpen
  \bibfield  {author} {\bibinfo {author} {\bibfnamefont {S.-Y.}\ \bibnamefont
  {Lin}}, \bibinfo {author} {\bibfnamefont {C.-H.}\ \bibnamefont {Chou}}, \
  and\ \bibinfo {author} {\bibfnamefont {B.~L.}\ \bibnamefont {Hu}},\ }\href
  {\doibase 10.1103/PhysRevD.91.084063} {\bibfield  {journal} {\bibinfo
  {journal} {Phys. Rev. D}\ }\textbf {\bibinfo {volume} {91}},\ \bibinfo
  {pages} {084063} (\bibinfo {year} {2015})}\BibitemShut {NoStop}%
\bibitem [{\citenamefont {Lin}(2014)}]{Lin2014projective}%
  \BibitemOpen
  \bibfield  {author} {\bibinfo {author} {\bibfnamefont {S.-Y.}\ \bibnamefont
  {Lin}},\ }\href {\doibase https://doi.org/10.1016/j.aop.2014.08.018}
  {\bibfield  {journal} {\bibinfo  {journal} {Annals of Physics}\ }\textbf
  {\bibinfo {volume} {351}},\ \bibinfo {pages} {773 } (\bibinfo {year}
  {2014})}\BibitemShut {NoStop}%
\bibitem [{\citenamefont {Hu}\ \emph {et~al.}(2012)\citenamefont {Hu},
  \citenamefont {Lin},\ and\ \citenamefont {Louko}}]{Hu2012review}%
  \BibitemOpen
  \bibfield  {author} {\bibinfo {author} {\bibfnamefont {B.~L.}\ \bibnamefont
  {Hu}}, \bibinfo {author} {\bibfnamefont {S.-Y.}\ \bibnamefont {Lin}}, \ and\
  \bibinfo {author} {\bibfnamefont {J.}~\bibnamefont {Louko}},\ }\href
  {http://stacks.iop.org/0264-9381/29/i=22/a=224005} {\bibfield  {journal}
  {\bibinfo  {journal} {Classical and Quantum Gravity}\ }\textbf {\bibinfo
  {volume} {29}},\ \bibinfo {pages} {224005} (\bibinfo {year}
  {2012})}\BibitemShut {NoStop}%
\bibitem [{\citenamefont {Sorkin}(1956)}]{sorkin1956}%
  \BibitemOpen
  \bibfield  {author} {\bibinfo {author} {\bibfnamefont {R.}~\bibnamefont
  {Sorkin}},\ }\enquote {\bibinfo {title} {Impossible measurements on quantum
  fields},}\ in\ \href {\doibase 10.1017/CBO9780511524653.024} {\emph {\bibinfo
  {booktitle} {Directions in General Relativity: Proceedings of the 1993
  International Symposium, Maryland: Papers in Honor of Dieter Brill}}},\
  Vol.~\bibinfo {volume} {2}\ (\bibinfo  {publisher} {Cambridge University
  Press},\ \bibinfo {year} {1956})\ pp.\ \bibinfo {pages}
  {293--305}\BibitemShut {NoStop}%
\bibitem [{\citenamefont {Benincasa}\ \emph {et~al.}(2014)\citenamefont
  {Benincasa}, \citenamefont {Borsten}, \citenamefont {Buck},\ and\
  \citenamefont {Dowker}}]{Benincasa2014projective}%
  \BibitemOpen
  \bibfield  {author} {\bibinfo {author} {\bibfnamefont {D.~M.~T.}\
  \bibnamefont {Benincasa}}, \bibinfo {author} {\bibfnamefont {L.}~\bibnamefont
  {Borsten}}, \bibinfo {author} {\bibfnamefont {M.}~\bibnamefont {Buck}}, \
  and\ \bibinfo {author} {\bibfnamefont {F.}~\bibnamefont {Dowker}},\ }\href
  {http://stacks.iop.org/0264-9381/31/i=7/a=075007} {\bibfield  {journal}
  {\bibinfo  {journal} {Classical and Quantum Gravity}\ }\textbf {\bibinfo
  {volume} {31}},\ \bibinfo {pages} {075007} (\bibinfo {year}
  {2014})}\BibitemShut {NoStop}%
\bibitem [{\citenamefont {{Dewitt}}(1979)}]{DeWitt1979}%
  \BibitemOpen
  \bibfield  {author} {\bibinfo {author} {\bibfnamefont {B.~S.}\ \bibnamefont
  {{Dewitt}}},\ }in\ \href@noop {} {\emph {\bibinfo {booktitle} {General
  Relativity: An Einstein centenary survey}}},\ \bibinfo {editor} {edited by\
  \bibinfo {editor} {\bibfnamefont {S.~W.}\ \bibnamefont {{Hawking}}}\ and\
  \bibinfo {editor} {\bibfnamefont {W.}~\bibnamefont {{Israel}}}}\ (\bibinfo
  {year} {1979})\ pp.\ \bibinfo {pages} {680--745}\BibitemShut {NoStop}%
\bibitem [{\citenamefont {Mart\'{\i}n-Mart\'{\i}nez}\ and\ \citenamefont
  {Rodriguez-Lopez}(2018)}]{Pablo2018rqo}%
  \BibitemOpen
  \bibfield  {author} {\bibinfo {author} {\bibfnamefont {E.}~\bibnamefont
  {Mart\'{\i}n-Mart\'{\i}nez}}\ and\ \bibinfo {author} {\bibfnamefont
  {P.}~\bibnamefont {Rodriguez-Lopez}},\ }\href {\doibase
  10.1103/PhysRevD.97.105026} {\bibfield  {journal} {\bibinfo  {journal} {Phys.
  Rev. D}\ }\textbf {\bibinfo {volume} {97}},\ \bibinfo {pages} {105026}
  (\bibinfo {year} {2018})}\BibitemShut {NoStop}%
\bibitem [{\citenamefont {Pozas-Kerstjens}\ and\ \citenamefont
  {Mart\'{\i}n-Mart\'{\i}nez}(2015)}]{pozas2015harvesting}%
  \BibitemOpen
  \bibfield  {author} {\bibinfo {author} {\bibfnamefont {A.}~\bibnamefont
  {Pozas-Kerstjens}}\ and\ \bibinfo {author} {\bibfnamefont {E.}~\bibnamefont
  {Mart\'{\i}n-Mart\'{\i}nez}},\ }\href {\doibase 10.1103/PhysRevD.92.064042}
  {\bibfield  {journal} {\bibinfo  {journal} {Phys. Rev. D}\ }\textbf {\bibinfo
  {volume} {92}},\ \bibinfo {pages} {064042} (\bibinfo {year}
  {2015})}\BibitemShut {NoStop}%
\bibitem [{\citenamefont {Takagi}(1986)}]{Takagi1986noise}%
  \BibitemOpen
  \bibfield  {author} {\bibinfo {author} {\bibfnamefont {S.}~\bibnamefont
  {Takagi}},\ }\href {\doibase 10.1143/PTP.88.1} {\bibfield  {journal}
  {\bibinfo  {journal} {Progress of Theoretical Physics Supplement}\ }\textbf
  {\bibinfo {volume} {88}},\ \bibinfo {pages} {1} (\bibinfo {year}
  {1986})}\BibitemShut {NoStop}%
\bibitem [{\citenamefont {Crispino}\ \emph {et~al.}(2008)\citenamefont
  {Crispino}, \citenamefont {Higuchi},\ and\ \citenamefont
  {Matsas}}]{Crispino2008review}%
  \BibitemOpen
  \bibfield  {author} {\bibinfo {author} {\bibfnamefont {L.~C.~B.}\
  \bibnamefont {Crispino}}, \bibinfo {author} {\bibfnamefont {A.}~\bibnamefont
  {Higuchi}}, \ and\ \bibinfo {author} {\bibfnamefont {G.~E.~A.}\ \bibnamefont
  {Matsas}},\ }\href {\doibase 10.1103/RevModPhys.80.787} {\bibfield  {journal}
  {\bibinfo  {journal} {Rev. Mod. Phys.}\ }\textbf {\bibinfo {volume} {80}},\
  \bibinfo {pages} {787} (\bibinfo {year} {2008})}\BibitemShut {NoStop}%
\bibitem [{\citenamefont
  {Mart\'{\i}n-Mart\'{\i}nez}(2015)}]{Causality2015Eduardo}%
  \BibitemOpen
  \bibfield  {author} {\bibinfo {author} {\bibfnamefont {E.}~\bibnamefont
  {Mart\'{\i}n-Mart\'{\i}nez}},\ }\href {\doibase 10.1103/PhysRevD.92.104019}
  {\bibfield  {journal} {\bibinfo  {journal} {Phys. Rev. D}\ }\textbf {\bibinfo
  {volume} {92}},\ \bibinfo {pages} {104019} (\bibinfo {year}
  {2015})}\BibitemShut {NoStop}%
\bibitem [{\citenamefont {Mart\'{\i}n-Mart\'{\i}nez}\ and\ \citenamefont
  {Louko}(2014)}]{Zeromode2014Edu}%
  \BibitemOpen
  \bibfield  {author} {\bibinfo {author} {\bibfnamefont {E.}~\bibnamefont
  {Mart\'{\i}n-Mart\'{\i}nez}}\ and\ \bibinfo {author} {\bibfnamefont
  {J.}~\bibnamefont {Louko}},\ }\href {\doibase 10.1103/PhysRevD.90.024015}
  {\bibfield  {journal} {\bibinfo  {journal} {Phys. Rev. D}\ }\textbf {\bibinfo
  {volume} {90}},\ \bibinfo {pages} {024015} (\bibinfo {year}
  {2014})}\BibitemShut {NoStop}%
\bibitem [{\citenamefont {Lin}\ \emph {et~al.}(2016)\citenamefont {Lin},
  \citenamefont {Chou},\ and\ \citenamefont {Hu}}]{Lin2016entangleCylin}%
  \BibitemOpen
  \bibfield  {author} {\bibinfo {author} {\bibfnamefont {S.-Y.}\ \bibnamefont
  {Lin}}, \bibinfo {author} {\bibfnamefont {C.-H.}\ \bibnamefont {Chou}}, \
  and\ \bibinfo {author} {\bibfnamefont {B.~L.}\ \bibnamefont {Hu}},\ }\href
  {\doibase 10.1007/JHEP03(2016)047} {\bibfield  {journal} {\bibinfo  {journal}
  {Journal of High Energy Physics}\ }\textbf {\bibinfo {volume} {2016}},\
  \bibinfo {pages} {47} (\bibinfo {year} {2016})}\BibitemShut {NoStop}%
\bibitem [{\citenamefont {Francesco}\ \emph {et~al.}(1996)\citenamefont
  {Francesco}, \citenamefont {Mathieu},\ and\ \citenamefont
  {Senechal}}]{francesco1996conformal}%
  \BibitemOpen
  \bibfield  {author} {\bibinfo {author} {\bibfnamefont {P.}~\bibnamefont
  {Francesco}}, \bibinfo {author} {\bibfnamefont {P.}~\bibnamefont {Mathieu}},
  \ and\ \bibinfo {author} {\bibfnamefont {D.}~\bibnamefont {Senechal}},\
  }\href {https://books.google.ca/books?id=mcMbswEACAAJ} {\emph {\bibinfo
  {title} {Conformal Field Theory}}}\ (\bibinfo  {publisher} {Island Press},\
  \bibinfo {year} {1996})\BibitemShut {NoStop}%
\bibitem [{\citenamefont {Allen}\ and\ \citenamefont
  {Folacci}(1987)}]{Allen1987deSitter}%
  \BibitemOpen
  \bibfield  {author} {\bibinfo {author} {\bibfnamefont {B.}~\bibnamefont
  {Allen}}\ and\ \bibinfo {author} {\bibfnamefont {A.}~\bibnamefont
  {Folacci}},\ }\href {\doibase 10.1103/PhysRevD.35.3771} {\bibfield  {journal}
  {\bibinfo  {journal} {Phys. Rev. D}\ }\textbf {\bibinfo {volume} {35}},\
  \bibinfo {pages} {3771} (\bibinfo {year} {1987})}\BibitemShut {NoStop}%
\bibitem [{\citenamefont {Kirsten}\ and\ \citenamefont
  {Garriga}(1993)}]{Kirsten1993deSitter}%
  \BibitemOpen
  \bibfield  {author} {\bibinfo {author} {\bibfnamefont {K.}~\bibnamefont
  {Kirsten}}\ and\ \bibinfo {author} {\bibfnamefont {J.}~\bibnamefont
  {Garriga}},\ }\href {\doibase 10.1103/PhysRevD.48.567} {\bibfield  {journal}
  {\bibinfo  {journal} {Phys. Rev. D}\ }\textbf {\bibinfo {volume} {48}},\
  \bibinfo {pages} {567} (\bibinfo {year} {1993})}\BibitemShut {NoStop}%
\bibitem [{\citenamefont {Page}\ and\ \citenamefont
  {Wu}(2012)}]{Page2012deSitter}%
  \BibitemOpen
  \bibfield  {author} {\bibinfo {author} {\bibfnamefont {D.~N.}\ \bibnamefont
  {Page}}\ and\ \bibinfo {author} {\bibfnamefont {X.}~\bibnamefont {Wu}},\
  }\href {http://stacks.iop.org/1475-7516/2012/i=11/a=051} {\bibfield
  {journal} {\bibinfo  {journal} {Journal of Cosmology and Astroparticle
  Physics}\ }\textbf {\bibinfo {volume} {2012}},\ \bibinfo {pages} {051}
  (\bibinfo {year} {2012})}\BibitemShut {NoStop}%
\bibitem [{\citenamefont {Bros}\ \emph {et~al.}(2010)\citenamefont {Bros},
  \citenamefont {Epstein},\ and\ \citenamefont {Moschella}}]{Bros2010}%
  \BibitemOpen
  \bibfield  {author} {\bibinfo {author} {\bibfnamefont {J.}~\bibnamefont
  {Bros}}, \bibinfo {author} {\bibfnamefont {H.}~\bibnamefont {Epstein}}, \
  and\ \bibinfo {author} {\bibfnamefont {U.}~\bibnamefont {Moschella}},\ }\href
  {\doibase 10.1007/s11005-010-0406-4} {\bibfield  {journal} {\bibinfo
  {journal} {Letters in Mathematical Physics}\ }\textbf {\bibinfo {volume}
  {93}},\ \bibinfo {pages} {203} (\bibinfo {year} {2010})}\BibitemShut
  {NoStop}%
\bibitem [{\citenamefont {Louko}\ and\ \citenamefont
  {Toussaint}(2016)}]{Louko2016fermionZM}%
  \BibitemOpen
  \bibfield  {author} {\bibinfo {author} {\bibfnamefont {J.}~\bibnamefont
  {Louko}}\ and\ \bibinfo {author} {\bibfnamefont {V.}~\bibnamefont
  {Toussaint}},\ }\href {\doibase 10.1103/PhysRevD.94.064027} {\bibfield
  {journal} {\bibinfo  {journal} {Phys. Rev. D}\ }\textbf {\bibinfo {volume}
  {94}},\ \bibinfo {pages} {064027} (\bibinfo {year} {2016})}\BibitemShut
  {NoStop}%
\bibitem [{\citenamefont {Yazdi}(2017)}]{Yazdi2017}%
  \BibitemOpen
  \bibfield  {author} {\bibinfo {author} {\bibfnamefont {Y.~K.}\ \bibnamefont
  {Yazdi}},\ }\href {\doibase 10.1007/JHEP04(2017)140} {\bibfield  {journal}
  {\bibinfo  {journal} {J. High Energ. Phys.}\ }\textbf {\bibinfo {volume}
  {2017}},\ \bibinfo {pages} {140} (\bibinfo {year} {2017})}\BibitemShut
  {NoStop}%
\bibitem [{\citenamefont {Robles}\ and\ \citenamefont
  {Rodr\'iguez-Laguna}(2017)}]{Robles2017thermometryQFT}%
  \BibitemOpen
  \bibfield  {author} {\bibinfo {author} {\bibfnamefont {S.}~\bibnamefont
  {Robles}}\ and\ \bibinfo {author} {\bibfnamefont {J.}~\bibnamefont
  {Rodr\'iguez-Laguna}},\ }\href
  {http://stacks.iop.org/1742-5468/2017/i=3/a=033105} {\bibfield  {journal}
  {\bibinfo  {journal} {Journal of Statistical Mechanics: Theory and
  Experiment}\ }\textbf {\bibinfo {volume} {2017}},\ \bibinfo {pages} {033105}
  (\bibinfo {year} {2017})}\BibitemShut {NoStop}%
\bibitem [{\citenamefont {Brown}(2013)}]{Brown2013Amplification}%
  \BibitemOpen
  \bibfield  {author} {\bibinfo {author} {\bibfnamefont {E.~G.}\ \bibnamefont
  {Brown}},\ }\href {\doibase 10.1103/PhysRevA.88.062336} {\bibfield  {journal}
  {\bibinfo  {journal} {Phys. Rev. A}\ }\textbf {\bibinfo {volume} {88}},\
  \bibinfo {pages} {062336} (\bibinfo {year} {2013})}\BibitemShut {NoStop}%
\bibitem [{\citenamefont {Brenna}\ \emph {et~al.}(2016)\citenamefont {Brenna},
  \citenamefont {Mann},\ and\ \citenamefont
  {Mart\'in-Mart\'inez}}]{Brenna2016antiUnruh}%
  \BibitemOpen
  \bibfield  {author} {\bibinfo {author} {\bibfnamefont {W.}~\bibnamefont
  {Brenna}}, \bibinfo {author} {\bibfnamefont {R.~B.}\ \bibnamefont {Mann}}, \
  and\ \bibinfo {author} {\bibfnamefont {E.}~\bibnamefont
  {Mart\'in-Mart\'inez}},\ }\href {\doibase
  https://doi.org/10.1016/j.physletb.2016.04.002} {\bibfield  {journal}
  {\bibinfo  {journal} {Physics Letters B}\ }\textbf {\bibinfo {volume}
  {757}},\ \bibinfo {pages} {307 } (\bibinfo {year} {2016})}\BibitemShut
  {NoStop}%
\bibitem [{\citenamefont {H\"ummer}\ \emph {et~al.}(2016)\citenamefont
  {H\"ummer}, \citenamefont {Mart\'{\i}n-Mart\'{\i}nez},\ and\ \citenamefont
  {Kempf}}]{Hummer2016bosonfermionZM}%
  \BibitemOpen
  \bibfield  {author} {\bibinfo {author} {\bibfnamefont {D.}~\bibnamefont
  {H\"ummer}}, \bibinfo {author} {\bibfnamefont {E.}~\bibnamefont
  {Mart\'{\i}n-Mart\'{\i}nez}}, \ and\ \bibinfo {author} {\bibfnamefont
  {A.}~\bibnamefont {Kempf}},\ }\href {\doibase 10.1103/PhysRevD.93.024019}
  {\bibfield  {journal} {\bibinfo  {journal} {Phys. Rev. D}\ }\textbf {\bibinfo
  {volume} {93}},\ \bibinfo {pages} {024019} (\bibinfo {year}
  {2016})}\BibitemShut {NoStop}%
\bibitem [{\citenamefont {Lorek}\ \emph {et~al.}(2014)\citenamefont {Lorek},
  \citenamefont {Pecak}, \citenamefont {Brown},\ and\ \citenamefont
  {Dragan}}]{Lorek2014tripartite}%
  \BibitemOpen
  \bibfield  {author} {\bibinfo {author} {\bibfnamefont {K.}~\bibnamefont
  {Lorek}}, \bibinfo {author} {\bibfnamefont {D.}~\bibnamefont {Pecak}},
  \bibinfo {author} {\bibfnamefont {E.~G.}\ \bibnamefont {Brown}}, \ and\
  \bibinfo {author} {\bibfnamefont {A.}~\bibnamefont {Dragan}},\ }\href
  {\doibase 10.1103/PhysRevA.90.032316} {\bibfield  {journal} {\bibinfo
  {journal} {Phys. Rev. A}\ }\textbf {\bibinfo {volume} {90}},\ \bibinfo
  {pages} {032316} (\bibinfo {year} {2014})}\BibitemShut {NoStop}%
\bibitem [{\citenamefont {Haag}\ and\ \citenamefont
  {Kastler}(1964)}]{HaagKastler1964algebraic}%
  \BibitemOpen
  \bibfield  {author} {\bibinfo {author} {\bibfnamefont {R.}~\bibnamefont
  {Haag}}\ and\ \bibinfo {author} {\bibfnamefont {D.}~\bibnamefont {Kastler}},\
  }\href {\doibase 10.1063/1.1704187} {\bibfield  {journal} {\bibinfo
  {journal} {Journal of Mathematical Physics}\ }\textbf {\bibinfo {volume}
  {5}},\ \bibinfo {pages} {848} (\bibinfo {year} {1964})}\BibitemShut {NoStop}%
\bibitem [{\citenamefont {Cliche}\ and\ \citenamefont
  {Kempf}(2010)}]{Cliche2010channel}%
  \BibitemOpen
  \bibfield  {author} {\bibinfo {author} {\bibfnamefont {M.}~\bibnamefont
  {Cliche}}\ and\ \bibinfo {author} {\bibfnamefont {A.}~\bibnamefont {Kempf}},\
  }\href {\doibase 10.1103/PhysRevA.81.012330} {\bibfield  {journal} {\bibinfo
  {journal} {Phys. Rev. A}\ }\textbf {\bibinfo {volume} {81}},\ \bibinfo
  {pages} {012330} (\bibinfo {year} {2010})}\BibitemShut {NoStop}%
\bibitem [{\citenamefont {Jonsson}\ \emph {et~al.}(2015)\citenamefont
  {Jonsson}, \citenamefont {Mart\'{\i}n-Mart\'{\i}nez},\ and\ \citenamefont
  {Kempf}}]{Jonsson2015infofree}%
  \BibitemOpen
  \bibfield  {author} {\bibinfo {author} {\bibfnamefont {R.~H.}\ \bibnamefont
  {Jonsson}}, \bibinfo {author} {\bibfnamefont {E.}~\bibnamefont
  {Mart\'{\i}n-Mart\'{\i}nez}}, \ and\ \bibinfo {author} {\bibfnamefont
  {A.}~\bibnamefont {Kempf}},\ }\href {\doibase 10.1103/PhysRevLett.114.110505}
  {\bibfield  {journal} {\bibinfo  {journal} {Phys. Rev. Lett.}\ }\textbf
  {\bibinfo {volume} {114}},\ \bibinfo {pages} {110505} (\bibinfo {year}
  {2015})}\BibitemShut {NoStop}%
\bibitem [{\citenamefont {Pozas-Kerstjens}\ \emph {et~al.}(2017)\citenamefont
  {Pozas-Kerstjens}, \citenamefont {Louko},\ and\ \citenamefont
  {Mart\'{\i}n-Mart\'{\i}nez}}]{PozasJorma}%
  \BibitemOpen
  \bibfield  {author} {\bibinfo {author} {\bibfnamefont {A.}~\bibnamefont
  {Pozas-Kerstjens}}, \bibinfo {author} {\bibfnamefont {J.}~\bibnamefont
  {Louko}}, \ and\ \bibinfo {author} {\bibfnamefont {E.}~\bibnamefont
  {Mart\'{\i}n-Mart\'{\i}nez}},\ }\href {\doibase 10.1103/PhysRevD.95.105009}
  {\bibfield  {journal} {\bibinfo  {journal} {Phys. Rev. D}\ }\textbf {\bibinfo
  {volume} {95}},\ \bibinfo {pages} {105009} (\bibinfo {year}
  {2017})}\BibitemShut {NoStop}%
\bibitem [{\citenamefont {Birrell}\ \emph {et~al.}(1984)\citenamefont
  {Birrell}, \citenamefont {Birrell},\ and\ \citenamefont
  {Davies}}]{birrell1984quantum}%
  \BibitemOpen
  \bibfield  {author} {\bibinfo {author} {\bibfnamefont {N.}~\bibnamefont
  {Birrell}}, \bibinfo {author} {\bibfnamefont {N.}~\bibnamefont {Birrell}}, \
  and\ \bibinfo {author} {\bibfnamefont {P.}~\bibnamefont {Davies}},\ }\href
  {https://books.google.ca/books?id=SEnaUnrqzrUC} {\emph {\bibinfo {title}
  {Quantum Fields in Curved Space}}},\ Cambridge Monographs on Mathematical
  Physics\ (\bibinfo  {publisher} {Cambridge University Press},\ \bibinfo
  {year} {1984})\BibitemShut {NoStop}%
\bibitem [{\citenamefont {Blasco}\ \emph {et~al.}(2015)\citenamefont {Blasco},
  \citenamefont {Garay}, \citenamefont {Mart\'{\i}n-Benito},\ and\
  \citenamefont {Mart\'{\i}n-Mart\'{\i}nez}}]{Blasco2015Huygens}%
  \BibitemOpen
  \bibfield  {author} {\bibinfo {author} {\bibfnamefont {A.}~\bibnamefont
  {Blasco}}, \bibinfo {author} {\bibfnamefont {L.~J.}\ \bibnamefont {Garay}},
  \bibinfo {author} {\bibfnamefont {M.}~\bibnamefont {Mart\'{\i}n-Benito}}, \
  and\ \bibinfo {author} {\bibfnamefont {E.}~\bibnamefont
  {Mart\'{\i}n-Mart\'{\i}nez}},\ }\href {\doibase
  10.1103/PhysRevLett.114.141103} {\bibfield  {journal} {\bibinfo  {journal}
  {Phys. Rev. Lett.}\ }\textbf {\bibinfo {volume} {114}},\ \bibinfo {pages}
  {141103} (\bibinfo {year} {2015})}\BibitemShut {NoStop}%
\bibitem [{\citenamefont {Blasco}\ \emph {et~al.}(2016)\citenamefont {Blasco},
  \citenamefont {Garay}, \citenamefont {Mart\'{\i}n-Benito},\ and\
  \citenamefont {Mart\'{\i}n-Mart\'{\i}nez}}]{Blasco2016broadcast}%
  \BibitemOpen
  \bibfield  {author} {\bibinfo {author} {\bibfnamefont {A.}~\bibnamefont
  {Blasco}}, \bibinfo {author} {\bibfnamefont {L.~J.}\ \bibnamefont {Garay}},
  \bibinfo {author} {\bibfnamefont {M.}~\bibnamefont {Mart\'{\i}n-Benito}}, \
  and\ \bibinfo {author} {\bibfnamefont {E.}~\bibnamefont
  {Mart\'{\i}n-Mart\'{\i}nez}},\ }\href {\doibase 10.1103/PhysRevD.93.024055}
  {\bibfield  {journal} {\bibinfo  {journal} {Phys. Rev. D}\ }\textbf {\bibinfo
  {volume} {93}},\ \bibinfo {pages} {024055} (\bibinfo {year}
  {2016})}\BibitemShut {NoStop}%
\bibitem [{\citenamefont {Sonego}\ and\ \citenamefont
  {Faraoni}(1992)}]{Sonego1992huygens}%
  \BibitemOpen
  \bibfield  {author} {\bibinfo {author} {\bibfnamefont {S.}~\bibnamefont
  {Sonego}}\ and\ \bibinfo {author} {\bibfnamefont {V.}~\bibnamefont
  {Faraoni}},\ }\href {\doibase 10.1063/1.529798} {\bibfield  {journal}
  {\bibinfo  {journal} {Journal of Mathematical Physics}\ }\textbf {\bibinfo
  {volume} {33}},\ \bibinfo {pages} {625} (\bibinfo {year} {1992})}\BibitemShut
  {NoStop}%
\bibitem [{\citenamefont {Faraoni}\ and\ \citenamefont
  {Gunzig}(1999)}]{Valerio1999tails}%
  \BibitemOpen
  \bibfield  {author} {\bibinfo {author} {\bibfnamefont {V.}~\bibnamefont
  {Faraoni}}\ and\ \bibinfo {author} {\bibfnamefont {E.}~\bibnamefont
  {Gunzig}},\ }\href {\doibase 10.1142/S021827189900016X} {\bibfield  {journal}
  {\bibinfo  {journal} {International Journal of Modern Physics D}\ }\textbf
  {\bibinfo {volume} {08}},\ \bibinfo {pages} {177} (\bibinfo {year}
  {1999})}\BibitemShut {NoStop}%
\end{thebibliography}%

\end{document}